\documentclass[pdflatex,sn-mathphys-num]{sn-jnl}

\usepackage{graphicx}
\usepackage{multirow}
\usepackage{amsmath,amssymb,amsfonts}
\usepackage{amsthm}
\usepackage{mathrsfs}
\usepackage[title]{appendix}
\usepackage{xcolor}
\usepackage{textcomp}
\usepackage{manyfoot}
\usepackage{booktabs}
\usepackage{algorithm}
\usepackage{algorithmicx}
\usepackage{algpseudocode}
\usepackage{listings}

\usepackage{rotating}
\usepackage{etoolbox}
\usepackage{url}
\usepackage{transparent}
\usepackage{setspace}
\usepackage{caption}
\usepackage{subcaption}
\usepackage{tikz}
\usetikzlibrary{calc,arrows.meta}
\usepackage{pgfplots}
\usepackage{bbding}
\usepackage{makecell}

\usepackage{cleveref}
\makeatletter
\crefname{@algorithm}{alg.}{algs.}
\Crefname{@algorithm}{Algorithm}{Algorithms}
\makeatother

\pgfplotsset{compat=1.18}

\theoremstyle{thmstyleone}

\theoremstyle{thmstyletwo}

\theoremstyle{thmstylethree}

\begin{document}

\title[Combinatorial maps for hierarchical splines]{Combinatorial maps for hierarchical splines}

\author*[1]{\fnm{Caleb B.} \sur{Goates} \orcid{https://orcid.org/0000-0001-5745-9902}}\email{calebgoates@gmail.com}
\author[1]{\fnm{Kendrick M.} \sur{Shepherd} \orcid{https://orcid.org/0000-0001-5562-4767}}
\author[2]{\fnm{Derek C.} \sur{Thomas} \orcid{https://orcid.org/0000-0003-4471-8762}}

\affil*[1]{\orgdiv{Department of Civil and Construction Engineering},
  \orgname{Brigham Young University},
  \orgaddress{\city{Provo}, \state{Utah}, \country{USA}}}

\affil[2]{\orgname{Coreform Inc}, \orgaddress{\city{Orem}, \state{Utah}, \country{USA}}}

\abstract{Hierarchical splines are an important part of multiscale and adaptive isogeometric analysis formulations.
The B\'ezier meshes of these splines are an essential part of their definition and of several important hierarchical spline algorithms, such as adaptive refinement and B\'ezier extraction.
Topological data associated with the B\'ezier mesh---such as adjacency information---can be used to improve the performance of many of these algorithms as well as downstream applications of the splines, but typical hierarchical spline formulations do not compute the topological data, storing instead just a list of elements.
In this work we present algorithms to build a performant topological data structure, namely the combinatorial map, to represent B\'ezier meshes of hierarchical splines over cubical cell complexes where the refinement levels have conforming B\'ezier meshes.
This includes hierarchical and truncated hierarchical B-splines, as well as subsets of other hierarchical spline formulations.
We show the performance characteristics of the construction algorithms of these hierarchical combinatorial maps, as well as an example use case, showing that the topological information can provide up to an order of magnitude reduction in computation time in downstream applications of the splines.}


\maketitle

\newcommand{\degree}{p}
\newcommand{\cmap}{\mathfrak{C}}
\newcommand{\dart}{\mathsf{d}}
\newcommand{\otherdart}{\mathsf{a}}
\newcommand{\numdarts}[1]{\left|#1\right|_\dart}
\newcommand{\numcells}[1]{\left|#1\right|}
\newcommand{\Dim}{n}
\newcommand{\NumRefinementLevels}{n_\ell}
\newcommand{\NumPatches}{n_p}
\newcommand{\LeafDarts}{\mathbf{B}_L}
\newcommand{\UnrefinedLeafDarts}{\mathbf{B}_U}
\newcommand{\NonLeafDarts}{\mathbf{B}_N}

\newcommand{\Rd}{\color{red}}
\newcommand{\Bd}{\color{blue}}

\section{Introduction}\label{sec:introduction}

B-splines play a critical role in computer-aided design and in computational analysis.
Initially finding widespread use in design, they more recently have been adopted for analysis with the introduction of isogeometric analysis \cite{Hughes:2005} where they offer higher accuracy per degree of freedom versus finite element analysis \cite{Sande:2020}, as well as better suitability for higher-order PDEs.
Because B-splines have a tensor product structure that is too restrictive for many scenarios, other splines have been introduced to isogeometric analysis which can provide similar analyses on less structured domains.
These include T-splines \cite{Bazilevs:2010}, U-splines \cite{Thomas:2022}, and LR-splines \cite{Dokken:2013}, all of which have received significant attention for their amenable behavior both in design and analysis.

As adaptive representation and resolution has become more important for computational data, hierarchical methods for isogeometric analysis have also appeared.
One tool for adaptive resolution that is finding increasing use in many domains is the hierarchical B-spline basis (HB-splines).
This basis was first published in 1988 for graphics applications \cite{Forsey:1988}, and was later adopted for isogeometric analysis (initially in \cite{Vuong:2011}).
Other hierarchical spline bases---such as truncated hierarchical B-splines \cite{Giannelli:2012}, hierarchical T-splines \cite{Evans:2015}, isogeometric spline forests \cite{Scott:2014}, and truncated hierarchical unstructured splines \cite{Wei:2022}---have also been developed for application to isogeometric analysis.
Many, if not most, implementations of isogeometric analysis take an elemental approach to integration, and the elements of these spline bases are defined using B\'ezier extraction \cite{Borden:2011,Scott:2011}.
These elements form a mesh with an associated topology.

Most existing papers on hierarchical B-splines and its variants do not treat in much detail the data structures that represent the splines.
Garau and Vasquez \cite{Garau:2018} present a summary of those that do, and in all of these (including \cite{Garau:2018} itself) the representation of the topology is ignored; most approaches simply store a set of elements that are found in the final hierarchical spline without any adjacency information.
While this is enough to define the basis itself, this representation does not support scalable topological operations that are useful for postprocessing and adaptivity.
For instance, to improve the stability and conditioning of the matrices in hierarchical analyses, hierarchical spline bases are often altered for admissibility, meaning that some neighboring elements of those that have been refined are also refined \cite{Buffa:2016}.
In addition, it may occur that refinement is desired near features, such as edges or vertices of the B\'ezier mesh.
These scenarios both benefit from a performant method to iterate neighborhoods of cells in the mesh.
Many postprocessing operations, such as extracting a watertight representation of level sets of functions defined on the hierarchical basis, also benefit from topological operations.

Outside isogeometric analysis, fields such as computer-aided design, scientific computing, computer graphics and computational geometry use topological data structures as a key component of their workflows.
Here the topology provides operations such as proximity checks, model refinement, model coarsening, and even simple representation of shape geometry.
Examples of such data structures include the half-edge data structure for polygonal surface modeling \cite{Muller:1978,Weiler:1985}, incidence graphs \cite{Logg:2009}, and boundary operators of chain complexes \cite{Shapero:2025}.
These data structures aim to be lightweight, navigable, and readily modifiable to enable the various operations necessary for their intended purposes.

A particularly general, efficient, and robust topological data structure is the combinatorial map, which provides topological operations on oriented $\Dim$-dimensional domains \cite{Edmonds:1960,Leinhardt:1988,Kraemer:2014}.
Combinatorial maps generalize the popular half-edge data structure of surfaces to higher dimensions and, when implemented appropriately, allow for rapid topological operations at relatively low storage expense.
In particular, a combinatorial map provides operations to navigate about the neighborhood of a particular cell in linear time with respect to the size of that neighborhood \cite{Kraemer:2014}.
Combinatorial maps have already found widespread use in computer graphics, computational geometry, geometric modeling, and mesh editing; Chapter 10 of \cite{Damiand:2015} provides an extensive review of relevant literature.

Generalizations of combinatorial maps to operate on subdivision surfaces and their higher-dimensional analogs have already been proposed to support computer-aided modeling for both simplicial and cubical cells \cite{Untereiner:2013}.
In addition, there are published algorithms that create combinatorial maps for quadtree and octree subdivision schemes \cite{Flechon:2014}.
Neither of these are straightforward to apply to HB-splines, however, especially when adaptive refinement schemes require repeatedly updating the underlying element mesh.

This work proposes construction algorithms for combinatorial maps representing the B\'ezier mesh of hierarchical B-splines and other hierarchical splines with cubical cells and conforming refinement level B\'ezier meshes.
An implementation of this data structure, developed as part of this work, is publicly available on GitHub \cite{sweeps_github}, and is leveraged therein for defining hierarchical spline spaces on unstructured domains.
In this work, emphasis will be placed on defining the combinatorial map data structure that underlies these splines.
For a detailed description of the splines themselves, we refer the reader to \cite{Vuong:2011,Giannelli:2012}.

The rest of this paper will proceed as follows.
\Cref{sec:background} will provide background information about the combinatorial map data structure, as well as motivating background about B-splines and hierarchical B-splines.
Next, \cref{sec:data_structures} will present the requirements for the nested combinatorial maps, give a low-memory implementation of a suitable combinatorial map for semi-structured meshes with cube-like cells, and give the algorithms necessary to create a hierarchical combinatorial map from the nested combinatorial maps.
\Cref{sec:results} will investigate the scaling of the algorithms as the number of elements grow and show the application of the combinatorial map to a relevant problem of calculating the level set of a function defined on the HB-splines, and \cref{sec:conclusions} will conclude the paper.

\section{Background}\label{sec:background}

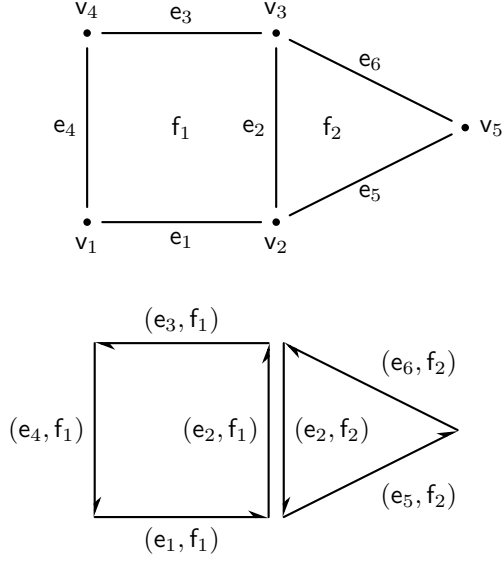
\begin{figure}
    \centering
\def\FC{black}
\def\BC{gray}

\begin{tikzpicture}[
  x={(2.5cm,0cm)},        
  y={(0cm,2.5cm)},        
  thick, line cap=round, line join=round
]

\pgfmathsetmacro{\EPS}{0.04}
\pgfmathsetmacro{\EP}{0.08}

\coordinate (A) at (0,0);
\coordinate (B) at (1,0);
\coordinate (C) at (1,1);
\coordinate (D) at (0,1);
\coordinate (E) at (2,0.5);
\coordinate (epsx) at (\EPS,0);
\coordinate (epsy) at (0,\EPS);
\coordinate (shift) at (0,-1.6);
\coordinate (epx) at (\EP,0);
\coordinate (epy) at (0,\EP);


\node[circle,fill,inner sep=1pt,label=below:$\mathsf{v}_1$] at (A) {};
\node[circle,fill,inner sep=1pt,label=below:$\mathsf{v}_2$] at (B) {};
\node[circle,fill,inner sep=1pt,label=above:$\mathsf{v}_3$] at (C) {};
\node[circle,fill,inner sep=1pt,label=above:$\mathsf{v}_4$] at (D) {};
\node[circle,fill,inner sep=1pt,label=right:$\mathsf{v}_5$] at (E) {};

\draw[-,\FC] ($(A) + (epx)$) -- ($(B)-(epx)$) node[midway,below] {$\mathsf{e}_1$};
\draw[-,\FC] ($(B) + (epy)$) -- ($(C)-(epy)$) node[midway,left] {$\mathsf{e}_2$};
\draw[-,\FC] ($(D) + (epx)$) -- ($(C)-(epx)$) node[midway,above] {$\mathsf{e}_3$};
\draw[-,\FC] ($(A) + (epy)$) -- ($(D)-(epy)$) node[midway,left] {$\mathsf{e}_4$};
\draw[-,\FC,shorten <=0.2cm,shorten >=0.2cm] ($(B)$) -- ($(E)$) node[midway,below] {$\mathsf{e}_5$};
\draw[-,\FC,shorten <=0.2cm,shorten >=0.2cm] ($(C)$) -- ($(E)$) node[midway,above] {$\mathsf{e}_6$};

\node at (0.5,0.5) {$\mathsf{f}_1$};
\node at (1.3,0.5) {$\mathsf{f}_2$};


\coordinate (Aleftdart) at ($(A) + (epsx) + (epsy) + (shift)$);
\coordinate (Bleftdart) at ($(B) - (epsx) + (epsy) + (shift)$);
\coordinate (Cleftdart) at ($(C) - (epsx) - (epsy) + (shift)$);
\coordinate (Dleftdart) at ($(D) + (epsx) - (epsy) + (shift)$);

\coordinate (Brightdart) at ($(B) + (epsx) + (epsy) + (shift)$);
\coordinate (Crightdart) at ($(C) + (epsx) - (epsy) + (shift)$);
\coordinate (Erightdart) at ($(E) - (epsx) + (shift)$);

\draw[-{Stealth[harpoon]},\FC] (Aleftdart) -- (Bleftdart) node[midway,below] {$(\mathsf{e}_1,\mathsf{f}_1)$};
\draw[-{Stealth[harpoon]},\FC] (Bleftdart) -- (Cleftdart) node[midway,left] {$(\mathsf{e}_2,\mathsf{f}_1)$};
\draw[-{Stealth[harpoon]},\FC] (Cleftdart) -- (Dleftdart) node[midway,above] {$(\mathsf{e}_3,\mathsf{f}_1)$};
\draw[-{Stealth[harpoon]},\FC] (Dleftdart) -- (Aleftdart) node[midway,left] {$(\mathsf{e}_4,\mathsf{f}_1)$};

\draw[-{Stealth[harpoon]},\FC] (Brightdart) -- (Erightdart) node[midway,below right] {$(\mathsf{e}_5,\mathsf{f}_2)$};
\draw[-{Stealth[harpoon]},\FC] (Erightdart) -- (Crightdart) node[midway,above right] {$(\mathsf{e}_6,\mathsf{f}_2)$};
\draw[-{Stealth[harpoon]},\FC] (Crightdart) -- (Brightdart) node[midway,right] {$(\mathsf{e}_2,\mathsf{f}_2)$};

\end{tikzpicture}
    \caption{A representation of darts in a combinatorial map as tuples, containing the cells of dimension greater than zero that can be represented by the dart. Inspired by Figs 1 and 2 of \cite{Kraemer:2014}.}
    \label{fig:dart_tuples}
\end{figure}

\subsection{Combinatorial Maps}

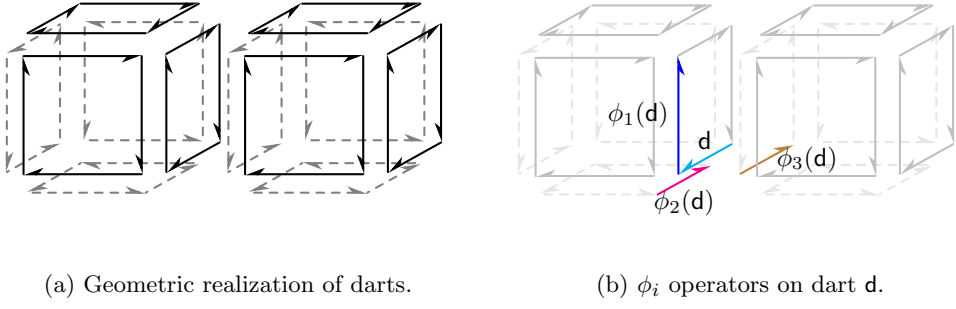
\begin{figure}
    \centering
    \begin{subfigure}{0.45\textwidth}
        \centering
\def\FC{black}
\def\BC{gray}

\resizebox{1\textwidth}{!}{%
\begin{tikzpicture}[
  x={(0.9cm,0.5cm)},        
  y={(2cm,0cm)},        
  z={(0cm,2cm)},    
  thick, line cap=round, line join=round
]

\pgfmathsetmacro{\EPS}{0.1}
\pgfmathsetmacro{\OFFSET}{0.4}

\coordinate (A) at (0,0,0);
\coordinate (B) at (1,0,0);
\coordinate (C) at (1,1,0);
\coordinate (D) at (0,1,0);
\coordinate (E) at (0,0,1);
\coordinate (F) at (1,0,1);
\coordinate (G) at (1,1,1);
\coordinate (H) at (0,1,1);
\coordinate (epsx) at (\EPS,0,0);
\coordinate (epsy) at (0,\EPS,0);
\coordinate (epsz) at (0,0,\EPS);
\coordinate (shift) at (0,1.5,0);

\coordinate (Abase) at ($(A) + (epsx) + (epsy) - \OFFSET*(epsz) $);
\coordinate (Bbase) at ($(B) - (epsx) + (epsy) - \OFFSET*(epsz)$);
\coordinate (Cbase) at ($(C) - (epsx) - (epsy) - \OFFSET*(epsz)$);
\coordinate (Dbase) at ($(D) + (epsx) - (epsy) - \OFFSET*(epsz)$);

\coordinate (Aleft) at ($(A) + (epsx) + (epsz) - \OFFSET*(epsy)$);
\coordinate (Bleft) at ($(B) - (epsx) + (epsz) - \OFFSET*(epsy)$);
\coordinate (Fleft) at ($(F) - (epsx) - (epsz) - \OFFSET*(epsy)$);
\coordinate (Eleft) at ($(E) + (epsx) - (epsz) - \OFFSET*(epsy)$);

\coordinate (Afront) at ($(A) + (epsy) + (epsz) + \OFFSET*(epsx)$);
\coordinate (Dfront) at ($(D) - (epsy) + (epsz) + \OFFSET*(epsx)$);
\coordinate (Hfront) at ($(H) - (epsy) - (epsz) + \OFFSET*(epsx)$);
\coordinate (Efront) at ($(E) + (epsy) - (epsz) + \OFFSET*(epsx)$);

\coordinate (Dright) at ($(D) + (epsx) + (epsz) + \OFFSET*(epsy)$);
\coordinate (Cright) at ($(C) - (epsx) + (epsz) + \OFFSET*(epsy)$);
\coordinate (Gright) at ($(G) - (epsx) - (epsz) + \OFFSET*(epsy)$);
\coordinate (Hright) at ($(H) + (epsx) - (epsz) + \OFFSET*(epsy)$);

\coordinate (Bback) at ($(B) + (epsy) + (epsz) - \OFFSET*(epsx)$);
\coordinate (Cback) at ($(C) - (epsy) + (epsz) - \OFFSET*(epsx)$);
\coordinate (Gback) at ($(G) - (epsy) - (epsz) - \OFFSET*(epsx)$);
\coordinate (Fback) at ($(F) + (epsy) - (epsz) - \OFFSET*(epsx)$);

\coordinate (Etop) at ($(E) + (epsx) + (epsy) + \OFFSET*(epsz)$);
\coordinate (Ftop) at ($(F) - (epsx) + (epsy)  + \OFFSET*(epsz)$);
\coordinate (Gtop) at ($(G) - (epsx) - (epsy)  + \OFFSET*(epsz)$);
\coordinate (Htop) at ($(H) + (epsx) - (epsy)  + \OFFSET*(epsz)$);

  \draw[{Stealth[harpoon]}-,\BC,dashed] (Abase) -- (Bbase);
  \draw[{Stealth[harpoon]}-,\BC,dashed] (Bbase) -- (Cbase);
  \draw[{Stealth[harpoon]}-,\BC,dashed] (Cbase) -- (Dbase);
  \draw[{Stealth[harpoon]}-,\BC,dashed] (Dbase) -- (Abase);
  
    \draw[{Stealth[harpoon]}-,\BC,dashed] (Bleft) -- (Aleft);
    \draw[{Stealth[harpoon]}-,\BC,dashed] (Fleft) -- (Bleft);
    \draw[{Stealth[harpoon]}-,\BC,dashed] (Eleft) -- (Fleft);
    \draw[{Stealth[harpoon]}-,\BC,dashed] (Aleft) -- (Eleft);

    \draw[{Stealth[harpoon]}-,\BC,dashed] (Cback) -- (Bback);
    \draw[{Stealth[harpoon]}-,\BC,dashed] (Bback) -- (Fback);
    \draw[{Stealth[harpoon]}-,\BC,dashed] (Fback) -- (Gback);
    \draw[{Stealth[harpoon]}-,\BC,dashed] (Gback) -- (Cback);

    \draw[{Stealth[harpoon,swap]}-,\FC] (Afront) -- (Dfront);
    \draw[{Stealth[harpoon,swap]}-,\FC] (Dfront) -- (Hfront);
    \draw[{Stealth[harpoon,swap]}-,\FC] (Hfront) -- (Efront);
    \draw[{Stealth[harpoon,swap]}-,\FC] (Efront) -- (Afront);

    \draw[{Stealth[harpoon,swap]}-,\FC] (Dright) -- (Cright);
    \draw[{Stealth[harpoon,swap]}-,\FC] (Cright) -- (Gright);
    \draw[{Stealth[harpoon,swap]}-,\FC] (Gright) -- (Hright);
    \draw[{Stealth[harpoon,swap]}-,\FC] (Hright) -- (Dright);

  \draw[{Stealth[harpoon,swap]}-,\FC] (Etop) -- (Htop);
  \draw[{Stealth[harpoon,swap]}-,\FC] (Htop) -- (Gtop);
  \draw[{Stealth[harpoon,swap]}-,\FC] (Gtop) -- (Ftop);
  \draw[{Stealth[harpoon,swap]}-,\FC] (Ftop) -- (Etop);


\coordinate (Abases) at ($(Abase) + (shift) $);
\coordinate (Bbases) at ($(Bbase) + (shift) $);
\coordinate (Cbases) at ($(Cbase) + (shift) $);
\coordinate (Dbases) at ($(Dbase) + (shift) $);

\coordinate (Alefts) at ($(Aleft) + (shift) $);
\coordinate (Blefts) at ($(Bleft) + (shift) $);
\coordinate (Flefts) at ($(Fleft) + (shift) $);
\coordinate (Elefts) at ($(Eleft) + (shift) $);

\coordinate (Afronts) at ($(Afront) + (shift) $);
\coordinate (Dfronts) at ($(Dfront) + (shift) $);
\coordinate (Hfronts) at ($(Hfront) + (shift) $);
\coordinate (Efronts) at ($(Efront) + (shift) $);

\coordinate (Drights) at ($(Dright) + (shift) $);
\coordinate (Crights) at ($(Cright) + (shift) $);
\coordinate (Grights) at ($(Gright) + (shift) $);
\coordinate (Hrights) at ($(Hright) + (shift) $);

\coordinate (Bbacks) at ($(Bback) + (shift) $);
\coordinate (Cbacks) at ($(Cback) + (shift) $);
\coordinate (Gbacks) at ($(Gback) + (shift) $);
\coordinate (Fbacks) at ($(Fback) + (shift) $);

\coordinate (Etops) at ($(Etop) + (shift) $);
\coordinate (Ftops) at ($(Ftop) + (shift) $);
\coordinate (Gtops) at ($(Gtop) + (shift) $);
\coordinate (Htops) at ($(Htop) + (shift) $);

  \draw[{Stealth[harpoon]}-,\BC,dashed] (Abases) -- (Bbases);
  \draw[{Stealth[harpoon]}-,\BC,dashed] (Bbases) -- (Cbases);
  \draw[{Stealth[harpoon]}-,\BC,dashed] (Cbases) -- (Dbases);
  \draw[{Stealth[harpoon]}-,\BC,dashed] (Dbases) -- (Abases);
  
    \draw[{Stealth[harpoon]}-,\BC,dashed] (Blefts) -- (Alefts);
    \draw[{Stealth[harpoon]}-,\BC,dashed] (Flefts) -- (Blefts);
    \draw[{Stealth[harpoon]}-,\BC,dashed] (Elefts) -- (Flefts);
    \draw[{Stealth[harpoon]}-,\BC,dashed] (Alefts) -- (Elefts);

    \draw[{Stealth[harpoon]}-,\BC,dashed] (Cbacks) -- (Bbacks);
    \draw[{Stealth[harpoon]}-,\BC,dashed] (Bbacks) -- (Fbacks);
    \draw[{Stealth[harpoon]}-,\BC,dashed] (Fbacks) -- (Gbacks);
    \draw[{Stealth[harpoon]}-,\BC,dashed] (Gbacks) -- (Cbacks);

    \draw[{Stealth[harpoon,swap]}-,\FC] (Afronts) -- (Dfronts);
    \draw[{Stealth[harpoon,swap]}-,\FC] (Dfronts) -- (Hfronts);
    \draw[{Stealth[harpoon,swap]}-,\FC] (Hfronts) -- (Efronts);
    \draw[{Stealth[harpoon,swap]}-,\FC] (Efronts) -- (Afronts);

    \draw[{Stealth[harpoon,swap]}-,\FC] (Drights) -- (Crights);
    \draw[{Stealth[harpoon,swap]}-,\FC] (Crights) -- (Grights);
    \draw[{Stealth[harpoon,swap]}-,\FC] (Grights) -- (Hrights);
    \draw[{Stealth[harpoon,swap]}-,\FC] (Hrights) -- (Drights);

  \draw[{Stealth[harpoon,swap]}-,\FC] (Etops) -- (Htops);
  \draw[{Stealth[harpoon,swap]}-,\FC] (Htops) -- (Gtops);
  \draw[{Stealth[harpoon,swap]}-,\FC] (Gtops) -- (Ftops);
  \draw[{Stealth[harpoon,swap]}-,\FC] (Ftops) -- (Etops);

\end{tikzpicture}
}
        \vspace{0.4cm}
        \caption{Geometric realization of darts.}
        \label{fig:dart3d_setup}
    \end{subfigure}
    \hspace{0.05\textwidth}
    \begin{subfigure}{0.45\textwidth}
        \centering
\def\FC{lightgray}
\def\BC{lightgray!40}

\resizebox{1\textwidth}{!}{%
\begin{tikzpicture}[
  x={(0.9cm,0.5cm)},        
  y={(2cm,0cm)},        
  z={(0cm,2cm)},    
  thick, line cap=round, line join=round
]

\pgfmathsetmacro{\EPS}{0.1}
\pgfmathsetmacro{\OFFSET}{0.4}

\coordinate (A) at (0,0,0);
\coordinate (B) at (1,0,0);
\coordinate (C) at (1,1,0);
\coordinate (D) at (0,1,0);
\coordinate (E) at (0,0,1);
\coordinate (F) at (1,0,1);
\coordinate (G) at (1,1,1);
\coordinate (H) at (0,1,1);
\coordinate (epsx) at (\EPS,0,0);
\coordinate (epsy) at (0,\EPS,0);
\coordinate (epsz) at (0,0,\EPS);
\coordinate (shift) at (0,1.5,0);

\coordinate (Abase) at ($(A) + (epsx) + (epsy) - \OFFSET*(epsz) $);
\coordinate (Bbase) at ($(B) - (epsx) + (epsy) - \OFFSET*(epsz)$);
\coordinate (Cbase) at ($(C) - (epsx) - (epsy) - \OFFSET*(epsz)$);
\coordinate (Dbase) at ($(D) + (epsx) - (epsy) - \OFFSET*(epsz)$);

\coordinate (Aleft) at ($(A) + (epsx) + (epsz) - \OFFSET*(epsy)$);
\coordinate (Bleft) at ($(B) - (epsx) + (epsz) - \OFFSET*(epsy)$);
\coordinate (Fleft) at ($(F) - (epsx) - (epsz) - \OFFSET*(epsy)$);
\coordinate (Eleft) at ($(E) + (epsx) - (epsz) - \OFFSET*(epsy)$);

\coordinate (Afront) at ($(A) + (epsy) + (epsz) + \OFFSET*(epsx)$);
\coordinate (Dfront) at ($(D) - (epsy) + (epsz) + \OFFSET*(epsx)$);
\coordinate (Hfront) at ($(H) - (epsy) - (epsz) + \OFFSET*(epsx)$);
\coordinate (Efront) at ($(E) + (epsy) - (epsz) + \OFFSET*(epsx)$);

\coordinate (Dright) at ($(D) + (epsx) + (epsz) + \OFFSET*(epsy)$);
\coordinate (Cright) at ($(C) - (epsx) + (epsz) + \OFFSET*(epsy)$);
\coordinate (Gright) at ($(G) - (epsx) - (epsz) + \OFFSET*(epsy)$);
\coordinate (Hright) at ($(H) + (epsx) - (epsz) + \OFFSET*(epsy)$);

\coordinate (Bback) at ($(B) + (epsy) + (epsz) - \OFFSET*(epsx)$);
\coordinate (Cback) at ($(C) - (epsy) + (epsz) - \OFFSET*(epsx)$);
\coordinate (Gback) at ($(G) - (epsy) - (epsz) - \OFFSET*(epsx)$);
\coordinate (Fback) at ($(F) + (epsy) - (epsz) - \OFFSET*(epsx)$);

\coordinate (Etop) at ($(E) + (epsx) + (epsy) + \OFFSET*(epsz)$);
\coordinate (Ftop) at ($(F) - (epsx) + (epsy)  + \OFFSET*(epsz)$);
\coordinate (Gtop) at ($(G) - (epsx) - (epsy)  + \OFFSET*(epsz)$);
\coordinate (Htop) at ($(H) + (epsx) - (epsy)  + \OFFSET*(epsz)$);

  \draw[{Stealth[harpoon]}-,\BC,dashed] (Abase) -- (Bbase);
  \draw[{Stealth[harpoon]}-,\BC,dashed] (Bbase) -- (Cbase);
  \draw[{Stealth[harpoon]}-,\BC,dashed] (Dbase) -- (Abase);
  
    \draw[{Stealth[harpoon]}-,\BC,dashed] (Bleft) -- (Aleft);
    \draw[{Stealth[harpoon]}-,\BC,dashed] (Fleft) -- (Bleft);
    \draw[{Stealth[harpoon]}-,\BC,dashed] (Eleft) -- (Fleft);
    \draw[{Stealth[harpoon]}-,\BC,dashed] (Aleft) -- (Eleft);

    \draw[{Stealth[harpoon]}-,\BC,dashed] (Cback) -- (Bback);
    \draw[{Stealth[harpoon]}-,\BC,dashed] (Bback) -- (Fback);
    \draw[{Stealth[harpoon]}-,\BC,dashed] (Fback) -- (Gback);
    \draw[{Stealth[harpoon]}-,\BC,dashed] (Gback) -- (Cback);

    \draw[{Stealth[harpoon,swap]}-,\FC] (Afront) -- (Dfront);
    \draw[{Stealth[harpoon,swap]}-,\FC] (Dfront) -- (Hfront);
    \draw[{Stealth[harpoon,swap]}-,\FC] (Hfront) -- (Efront);
    \draw[{Stealth[harpoon,swap]}-,\FC] (Efront) -- (Afront);

    \draw[{Stealth[harpoon,swap]}-,\FC] (Gright) -- (Hright);
    \draw[{Stealth[harpoon,swap]}-,\FC] (Cright) -- (Gright);

  \draw[{Stealth[harpoon,swap]}-,\FC] (Etop) -- (Htop);
  \draw[{Stealth[harpoon,swap]}-,\FC] (Htop) -- (Gtop);
  \draw[{Stealth[harpoon,swap]}-,\FC] (Gtop) -- (Ftop);
  \draw[{Stealth[harpoon,swap]}-,\FC] (Ftop) -- (Etop);

  \draw[{Stealth[harpoon,swap]}-,cyan] (Dright) -- (Cright) node[midway, above] {\color{black} $\mathsf{d}$};
  \draw[{Stealth[harpoon]}-,magenta] (Cbase) -- (Dbase) node[midway,below] {\color{black} $\phi_2(\mathsf{d})$};


\coordinate (Abases) at ($(Abase) + (shift) $);
\coordinate (Bbases) at ($(Bbase) + (shift) $);
\coordinate (Cbases) at ($(Cbase) + (shift) $);
\coordinate (Dbases) at ($(Dbase) + (shift) $);

\coordinate (Alefts) at ($(Aleft) + (shift) $);
\coordinate (Blefts) at ($(Bleft) + (shift) $);
\coordinate (Flefts) at ($(Fleft) + (shift) $);
\coordinate (Elefts) at ($(Eleft) + (shift) $);

\coordinate (Afronts) at ($(Afront) + (shift) $);
\coordinate (Dfronts) at ($(Dfront) + (shift) $);
\coordinate (Hfronts) at ($(Hfront) + (shift) $);
\coordinate (Efronts) at ($(Efront) + (shift) $);

\coordinate (Drights) at ($(Dright) + (shift) $);
\coordinate (Crights) at ($(Cright) + (shift) $);
\coordinate (Grights) at ($(Gright) + (shift) $);
\coordinate (Hrights) at ($(Hright) + (shift) $);

\coordinate (Bbacks) at ($(Bback) + (shift) $);
\coordinate (Cbacks) at ($(Cback) + (shift) $);
\coordinate (Gbacks) at ($(Gback) + (shift) $);
\coordinate (Fbacks) at ($(Fback) + (shift) $);

\coordinate (Etops) at ($(Etop) + (shift) $);
\coordinate (Ftops) at ($(Ftop) + (shift) $);
\coordinate (Gtops) at ($(Gtop) + (shift) $);
\coordinate (Htops) at ($(Htop) + (shift) $);

  \draw[{Stealth[harpoon]}-,\BC,dashed] (Abases) -- (Bbases);
  \draw[{Stealth[harpoon]}-,\BC,dashed] (Bbases) -- (Cbases);
  \draw[{Stealth[harpoon]}-,\BC,dashed] (Cbases) -- (Dbases);
  \draw[{Stealth[harpoon]}-,\BC,dashed] (Dbases) -- (Abases);
  
    \draw[{Stealth[harpoon]}-,\BC,dashed] (Flefts) -- (Blefts);
    \draw[{Stealth[harpoon]}-,\BC,dashed] (Elefts) -- (Flefts);
    \draw[{Stealth[harpoon]}-,\BC,dashed] (Alefts) -- (Elefts);

    \draw[{Stealth[harpoon]}-,\BC,dashed] (Cbacks) -- (Bbacks);
    \draw[{Stealth[harpoon]}-,\BC,dashed] (Bbacks) -- (Fbacks);
    \draw[{Stealth[harpoon]}-,\BC,dashed] (Fbacks) -- (Gbacks);
    \draw[{Stealth[harpoon]}-,\BC,dashed] (Gbacks) -- (Cbacks);

    \draw[{Stealth[harpoon,swap]}-,\FC] (Afronts) -- (Dfronts);
    \draw[{Stealth[harpoon,swap]}-,\FC] (Dfronts) -- (Hfronts);
    \draw[{Stealth[harpoon,swap]}-,\FC] (Hfronts) -- (Efronts);

    \draw[{Stealth[harpoon,swap]}-,\FC] (Drights) -- (Crights);
    \draw[{Stealth[harpoon,swap]}-,\FC] (Crights) -- (Grights);
    \draw[{Stealth[harpoon,swap]}-,\FC] (Grights) -- (Hrights);
    \draw[{Stealth[harpoon,swap]}-,\FC] (Hrights) -- (Drights);

  \draw[{Stealth[harpoon,swap]}-,\FC] (Etops) -- (Htops);
  \draw[{Stealth[harpoon,swap]}-,\FC] (Htops) -- (Gtops);
  \draw[{Stealth[harpoon,swap]}-,\FC] (Gtops) -- (Ftops);
  \draw[{Stealth[harpoon,swap]}-,\FC] (Ftops) -- (Etops);

    \draw[{Stealth[harpoon]}-,brown] (Blefts) -- (Alefts) node[midway,right] {\color{black} $\phi_3(\mathsf{d})$};
    \draw[{Stealth[harpoon,swap]}-,\FC] (Efronts) -- (Afronts);
    \draw[{Stealth[harpoon,swap]}-,blue] (Hright) -- (Dright) node[midway,left] {\color{black} $\phi_1(\mathsf{d})$};

\end{tikzpicture}
}
        \caption{$\phi_i$ operators on dart $\mathsf{d}$.}
        \label{fig:dart3d_phis}
    \end{subfigure}
    \caption{A pictorial representation of a combinatorial map on a domain composed of two hexahedra sharing a common face. In \cref{fig:dart3d_setup}, a geometric representation of each dart (associated with an edge, face, and volume of the domain) is shown for an exploded representation of the domain. Given a dart associated with the shared face of both volumes, $\mathsf{d}$ (shown in cyan), operators $\phi_1$, $\phi_2$, and $\phi_3$ are depicted in \cref{fig:dart3d_phis} to show how the domain can be operated on, with $\phi_1(\mathsf{d})$ yielding the dart in blue, $\phi_2(\mathsf{d})$ yielding the dart in magenta, and $\phi_3(\mathsf{d})$ yielding the brown dart. Note that the $\phi_i$ operator shifts both the originating vertex and the $i$-dimensional cell associated with the cell while leaving all other cells constant.}
    \label{fig:dart3d}
\end{figure}

A combinatorial map $\cmap$ is a representation of a mesh as a set of abstract entities, called darts.
Each dart roughly corresponds to a tuple of cells from dimension 1 to $\Dim$, and the combinatorial map includes relations to traverse the darts.
\Cref{fig:dart_tuples} shows a simple two-dimensional mesh consisting of one quadrilateral face and one triangular face with all the cells labeled.
Below that is shown a geometric representation of the combinatorial map corresponding to the mesh with each dart labeled with its cell tuple.
In a two-dimensional combinatorial map, each dart (represented as a half arrow in the figure) corresponds to an edge-face tuple, and there is a dart for each combination of edge and face which are adjacent to each other.
The relations to traverse the darts are denoted as $\phi_i:\operatorname{darts}(\cmap)\rightarrow\operatorname{darts}(\cmap)\cup\emptyset$, where $i\in\{-1,1,\ldots,\Dim\}$ and $\operatorname{darts}(\cmap)$ indicates the set of all darts in $\cmap$.
In particular, the $\phi_1$ operation returns the dart on the next edge while keeping all other elements of the tuple as they were, the $\phi_{-1}$ operation returns the dart on the previous edge, and the $\phi_k$ for $1<k\leq\Dim$ traverses to the dart with the $k$-cell in the tuple changed to the only other $k$-cell that is adjacent to the other cells in the tuple.
Example $\phi_i$ of a dart $\mathsf{d}$ in a 3-dimensional combinatorial map are shown in \cref{fig:dart3d_phis}.
Which edge is next or previous is decided by convention for a given implementation, but the decision that is made defines a direction of sorts that a dart is pointing in, denoted by the direction of the arrows in \cref{fig:dart_tuples,fig:dart3d}.
In this work we choose the $\phi_1$ direction such that for surfaces it follows the right-hand rule subject to the surface normal, and for volumes it follows the right-hand rule subject to a face normal pointing into the interior of the volumetric cell.
By analogy to an arrow, we can refer to the vertices adjacent to a dart as being at the head or the tail of the dart: the vertex the arrow is pointing toward is at the head, and the vertex the arrow is pointing away from is at the tail.
On boundaries, there may be no suitable dart for the $\phi_\Dim$ operation to return; some implementations handle this by returning nothing, while others treat each connected boundary component of the domain as an $\Dim$-cell of its own.
In this work we allow $\phi_\Dim$ operations to return nothing, which is why the range of the $\phi_i$ operator includes $\emptyset$.
The combinatorial map as treated here requires that the meshes it represents be manifold---each $(\Dim-1)$-cell neighbors at most two $\Dim$-cells---and orientable.
Extensions exist to remove those restrictions, but we do not treat those in this paper.
We denote the number of $\Dim$-cells in $\cmap$ as $\numcells{\cmap}$ and the number of darts in $\cmap$ as $\numdarts{\cmap}$, and we use the term \textit{elements of $\cmap$} to refer to $\Dim$-cells.
When necessary, we will also indicate that a $\phi_i$ operation is performed in a specific combinatorial map $\cmap$ using the notation $\phi_i[\cmap](\dart)$.

The tuples that the darts represent are a tool for understanding darts; they are not explicitly stored.
Instead, each dart is represented as an index, or an unsigned integer.
Cells in combinatorial maps may be indicated by a cell dimension and any one of the darts that would contain that cell in its tuple.
For vertices, any dart whose tail lies at the vertex may be used as a representation of the vertex when paired with a cell dimension of zero.
We denote the set of all darts of a cell $\mathsf{c}$ as $\operatorname{dartsOf}(\mathsf{c})$ and the dart that is used as a representation of the cell as $\mathsf{c}.\mathrm{dart}()$.
We also define operators $\operatorname{edge}(\cdot)$ and $\operatorname{face}(\cdot)$ that operate on a dart to give the 1d or 2d cells defined by that dart.

Combinatorial maps contain only the topological description of the mesh.
The geometric representation of the mesh, and indeed any data desired, can be layered on top of the combinatorial map by assigning a pointer to the data that corresponds to a given cell with each dart that represents the cell.

There are convenient algorithms using the $\phi$ operations to find all adjacencies of any given cell, traverse neighborhoods, etc. with linear time complexity with respect to the size of the neighborhood being traversed.
An overview of many of these algorithms can be found in Chapter 6 of \cite{Damiand:2015}.
In particular here we will use the operator $\operatorname{Adj}^k(\cdot)$, which gives all the $k$-cells adjacent to a given cell.

In practice, a combinatorial map with $\numdarts{\cmap}$ darts can be represented as $\Dim+1$ arrays of length $\numdarts{\cmap}$, where each array stores at index $d$ one of the $\phi$ operations of the dart with index $d$ \cite{Kraemer:2014}.
Additional attributes, such as cell indices, can be stored in additional arrays.
Other implementations designed for specific applications use the structure of the mesh (e.g. an all tetrahedral mesh) to store the combinatorial map more compactly by implicitly representing some $\phi$ operations \cite{Feng:2013}.
Our implementation uses the latter approach to define combinatorial maps for global refinement levels from which the hierarchical combinatorial map is defined.

\subsection{B-Splines}\label{sec:hier_bsplines}

A one-dimensional B-spline basis is defined by a polynomial degree $\degree$ and a knot vector $\Xi=\{\zeta_0,\ldots,\zeta_m\}$, which is a non-decreasing list of knot values $\zeta_i$.
Given this data, $m-p$ B-splines $\left\{B_i^\degree[\Xi](\xi)\right\}_{i=0}^{m-p-1}$ can be constructed using the Cox--de Boor recurrence formula \cite[Section 2.2]{piegl_nurbs}.
These span a space $\mathcal{B}^\degree[\Xi]$ of piecewise polynomial functions, called the B-spline space.

The B-spline basis can be extended to $\Dim$ dimensions for degrees $\left\{\degree_k\right\}_{k=0}^{\Dim-1}$ and knot vectors $\left\{\Xi_k\right\}_{k=0}^{\Dim-1}$ by defining the $\Dim$-dimensional B-splines as
\begin{equation*}
B_{i_0,\ldots,i_{\Dim-1}}^{\degree_0,\ldots,\degree_{\Dim-1}}[\Xi_0,\ldots,\Xi_{\Dim-1}](\xi_0,\ldots,\xi_{\Dim-1})=
\prod_{k=0}^{\Dim-1} B_{i_k}^{\degree_k}[\Xi_k](\xi_k).
\end{equation*}
These higher-dimensional B-splines span an associated B-spline space with parametric dimension $\Dim$, which we will denote as $\mathcal{B}^{\degree_0,\ldots,\degree_{\Dim-1}}[\Xi_0,\ldots,\Xi_{\Dim-1}]$.
For brevity, we will let $\mathbf{I}=i_0,\ldots,i_{\Dim-1}$, $\mathbf{\degree}=\degree_0,\ldots,\degree_{\Dim-1}$, $\mathbf{\Xi}=\Xi_0,\ldots,\Xi_{\Dim-1}$, and $\boldsymbol{\xi}=\xi_0,\ldots,\xi_{\Dim-1}$, and write tensor product B-splines as $B_\mathbf{I}^\mathbf{\degree}[\mathbf{\Xi}](\boldsymbol{\xi})$, and the associated space as $\mathcal{B}^\mathbf{\degree}[\mathbf{\Xi}]$.
The domain of definition of these splines is called the parametric domain, which we denote as $\hat\Omega$.
B-splines defined in this way can be extended to any domain $\Omega$ which is homeomorphic to an $\Dim$-cube with a mapping $F:\hat\Omega\rightarrow\Omega$ as
$B_\mathbf{I}^\mathbf{\degree}[\mathbf{\Xi}](\mathbf{x}):=B_\mathbf{I}^\mathbf{\degree}[\mathbf{\Xi}](F^{-1}(\mathbf{x}))$.

We can also define B-splines over more exotic physical domains which are decomposed into $n_p$ domains $\Omega^i$ homeomorphic to an $\Dim$-cube using a multi-patch formulation.
This is done over a $\Dim$-dimensional domain $\mathcal{M}$ with a decomposition $\left\{(\Omega^i,F_i)\right\}_{i=0}^{n_p-1}$ that satisfies the following requirements:
\begin{itemize}
\item $\bigcup_i \Omega^i = \mathcal{M}$.
\item $\Omega^i\cap\Omega^j$ for $i\neq j$ is either empty, or of dimension $\Dim-1$ or lower.
\item Each $F_i$ homeomorphically maps from a rectangular domain $\hat{\Omega}^i$ to $\Omega^i$.
\item When $\Omega^i\cap\Omega^j\neq\emptyset$, the transition map $\psi_{ij} = F_i^{-1}\circ F_j$ is either the identity, or some scaling, flip, or rotation.
\end{itemize}
We then define a set of B-splines over each $\Omega^i$ as in the single patch case.
If $C^0$ continuity between $\Omega^i$ and $\Omega^j$ is desired when $\Omega^i\cap\Omega^j\neq\emptyset$, then any B-spline $B_\mathbf{I}[\mathbf{\Xi}_i]$ defined over $\Omega^i$ which is nonzero on $\Omega^i\cap\Omega^j$ must have a $B_{\mathbf{I}'}[\mathbf{\Xi}_j]$ defined over $\Omega^j$ such that $B_\mathbf{I}[\mathbf{\Xi}_i]\circ\psi_{ij}=B_{\mathbf{I}'}[\mathbf{\Xi}_j]$.
Geometric continuity is enforced by equating the control points of these matching B-splines.
Higher continuity on portions of $\Omega^i\cap\Omega^j$ is also possible in some spline formulations, such as T-splines or U-splines.

For each B-spline space mentioned above, there exists an underlying mesh formed by the non-zero knot intervals in the knot vectors.
This mesh is the main entity that we treat in this work, and is called a B\'ezier mesh \cite{Borden:2011}.
The algorithm that we present in \cref{sec:hierarchical_algorithm} applies to a hierarchy of any bases whose B\'ezier meshes are conforming, i.e., have no T-junctions.
While this includes splines over non-cubelike cells, such as hierarchical Powell-Sabin splines \cite{Maes:2006}, our data structures are made feasible by implicitly representing the B\'ezier meshes in the hierarchy using the structure available in all-hex and all-quad meshes, and their application to B\'ezier meshes including other cells would likely incur a high memory cost.

Hierarchical B-splines are created by selecting basis functions from a series of $n_\ell$ nested B-spline spaces over the physical domain $\Omega$,
\begin{equation*}
    \mathcal{B}_0\subset\mathcal{B}_1\subset\ldots\subset\mathcal{B}_{n_\ell},
\end{equation*}
where we have suppressed the superscript representing the degree.
With the nested spaces there is also a set of nested domains
\begin{equation*}
    \Omega_0\supseteq\Omega_1\supseteq\ldots\supseteq\Omega_{n_\ell}
\end{equation*}
which determine the B-splines chosen from the $k$th level as those who are zero on $\Omega\setminus\Omega_k$ but are not zero on $\Omega\setminus\Omega_{k+1}$.
This final collection of B-splines spans the hierarchical B-spline space $\mathcal{H}$.
Truncated hierarchical B-splines \cite{Giannelli:2012} share the same B\'ezier mesh as hierarchical B-splines, and the combinatorial map data structures presented herein are equally applicable to them, as well as other hierarchical bases.
For brevity, we will not give the formulation of these other hierarchical bases in this work.
This work also requires that each $\Omega_k$ for $k>0$ be the union of the closure of elements from the B\'ezier mesh on level $k-1$.
The elements in the hierarchical B\'ezier mesh are taken from the B\'ezier meshes of each nested spline space and the elements taken from the $k$th refinement level are those contained within $\Omega_k$ but not within $\Omega_{k+1}$.
We call these elements leaf elements.

Note that while the B\'ezier meshes of the B-spline spaces in the hierarchy must be conforming, the final hierarchical B\'ezier mesh is not conforming and will include T-junctions in all but the most trivial of cases.

\section{Data Structures}\label{sec:data_structures}

While the main contribution of this work is the two algorithms that define a hierarchical combinatorial map at the end of this section, the algorithms themselves are made feasible by using a set of implicitly represented combinatorial maps for the B\'ezier meshes of the nested spline spaces.
\Cref{alg:hierleafdarts,alg:hierphi1} can be used without the implicit refinement level maps, but at a much higher memory cost that may make them infeasible.
In this section we will first present, therefore, algorithms needed to define a low-memory combinatorial map that can represent semi-structured quadrilateral or hexahedral cell complexes with no T-junctions.
Afterward we will present the algorithms that can take these low-memory global refinement levels and build the hierarchical combinatorial map from them.
In the following sections where it would become awkward to repeatedly write combinatorial map, we use instead the abbreviation C-map.

\subsection{Refinement Levels}

We begin by defining one-dimensional combinatorial maps, and build higher-dimensional combinatorial maps as tensor products of 1d maps.

\subsubsection{One-dimensional}
For a 1d combinatorial map $\cmap_{\mathrm{1d}}$, the only operations needed are $\phi_1$ and $\phi_{-1}$.
Given the number of edges $\numcells{\cmap_{\mathrm{1d}}}$ and whether $\cmap_{\mathrm{1d}}$ is periodic, we can define the $\phi_1$ operation as
\begin{equation}
    \phi_1(\dart)=\begin{cases}
    \dart+1&\dart<\numcells{\cmap_{\mathrm{1d}}}-1\\
    0&\dart=\numcells{\cmap_{\mathrm{1d}}}-1 \text{ and } \cmap_{\mathrm{1d}} \text{ is periodic}\\
    \emptyset&\text{otherwise.}
    \end{cases}
\end{equation}
The $\phi_{-1}$ operation can be defined as $\phi_{-1}(\dart)=\phi_1^{-1}(\dart)$ where that inverse exists, and $\emptyset$ otherwise.

\subsubsection{Tensor-product}

Given an $(\Dim-1)$-dimensional ($\Dim=2,3$) C-map $\cmap_{\mathrm{base}}$ and a one-dimensional C-map $\cmap_{\mathrm{line}}$, we can define an $\Dim$-dimensional C-map $\cmap_{\mathrm{tp}}$ with dart and cell counts
\begin{align*}
&\numdarts{\cmap_{\mathrm{tp}}} = 2\Dim\numdarts{\cmap_{\mathrm{base}}}\numdarts{\cmap_{\mathrm{line}}}\\
&\numcells{\cmap_{\mathrm{tp}}} = \numcells{\cmap_{\mathrm{base}}}\numcells{\cmap_{\mathrm{line}}}.
\end{align*}
We call $\cmap_{\mathrm{tp}}$ a \textit{tensor product} of $\cmap_{\mathrm{base}}$ and $\cmap_{\mathrm{line}}$ and write the relationship as $\cmap_{\mathrm{tp}} = \cmap_{\mathrm{base}}\times\cmap_{\mathrm{line}}$.
Every dart $\dart_{\mathrm{tp}}\in\cmap_{\mathrm{tp}}$ corresponds to a tuple $(\dart_{\mathrm{base}}, \dart_\mathrm{line}, a)$, where $\dart_{\mathrm{base}}\in\cmap_\mathrm{base}$, $\dart_{\mathrm{line}}\in\cmap_\mathrm{line}$, and $a\in\{0,1,...,2\Dim-1\}$, and its index can be calculated as
\begin{align}
    \begin{split}
\dart_{\mathrm{tp}}&=\textrm{flatten}(\dart_{\mathrm{base}}, \dart_\mathrm{line}, a)\\
&= a+2\Dim\dart_{\mathrm{base}}+2\Dim\numdarts{\cmap_{\mathrm{base}}}\dart_\mathrm{line}.
    \end{split}
\end{align}
We also denote the reverse operation as
\begin{equation}
(\dart_\mathrm{base},\dart_\mathrm{line},a)=\textrm{unflatten}(\dart_\mathrm{tp}),
\end{equation}
which can be calculated using integer division and the modulo operator.

\begin{figure}
	\centering
\def\FC{gray}
\def\BC{lightgray}

\resizebox{0.3\textwidth}{!}{%
\begin{tikzpicture}[
  x={(0.9cm,0.5cm)},        
  y={(2cm,0cm)},        
  z={(0cm,2cm)},    
  thick, line cap=round, line join=round
]

\pgfmathsetmacro{\EPS}{0.1}
\pgfmathsetmacro{\OFFSET}{0.4}

\coordinate (A) at (0,0,0);
\coordinate (B) at (1,0,0);
\coordinate (C) at (1,1,0);
\coordinate (D) at (0,1,0);
\coordinate (E) at (0,0,1);
\coordinate (F) at (1,0,1);
\coordinate (G) at (1,1,1);
\coordinate (H) at (0,1,1);
\coordinate (epsx) at (\EPS,0,0);
\coordinate (epsy) at (0,\EPS,0);
\coordinate (epsz) at (0,0,\EPS);
\coordinate (shift) at (0,1.5,0);

\coordinate (Abase) at ($(A) + (epsx) + (epsy) - \OFFSET*(epsz) $);
\coordinate (Bbase) at ($(B) - (epsx) + (epsy) - \OFFSET*(epsz)$);
\coordinate (Cbase) at ($(C) - (epsx) - (epsy) - \OFFSET*(epsz)$);
\coordinate (Dbase) at ($(D) + (epsx) - (epsy) - \OFFSET*(epsz)$);

\coordinate (Aleft) at ($(A) + (epsx) + (epsz) - \OFFSET*(epsy)$);
\coordinate (Bleft) at ($(B) - (epsx) + (epsz) - \OFFSET*(epsy)$);
\coordinate (Fleft) at ($(F) - (epsx) - (epsz) - \OFFSET*(epsy)$);
\coordinate (Eleft) at ($(E) + (epsx) - (epsz) - \OFFSET*(epsy)$);

\coordinate (Afront) at ($(A) + (epsy) + (epsz) + \OFFSET*(epsx)$);
\coordinate (Dfront) at ($(D) - (epsy) + (epsz) + \OFFSET*(epsx)$);
\coordinate (Hfront) at ($(H) - (epsy) - (epsz) + \OFFSET*(epsx)$);
\coordinate (Efront) at ($(E) + (epsy) - (epsz) + \OFFSET*(epsx)$);

\coordinate (Dright) at ($(D) + (epsx) + (epsz) + \OFFSET*(epsy)$);
\coordinate (Cright) at ($(C) - (epsx) + (epsz) + \OFFSET*(epsy)$);
\coordinate (Gright) at ($(G) - (epsx) - (epsz) + \OFFSET*(epsy)$);
\coordinate (Hright) at ($(H) + (epsx) - (epsz) + \OFFSET*(epsy)$);

\coordinate (Bback) at ($(B) + (epsy) + (epsz) - \OFFSET*(epsx)$);
\coordinate (Cback) at ($(C) - (epsy) + (epsz) - \OFFSET*(epsx)$);
\coordinate (Gback) at ($(G) - (epsy) - (epsz) - \OFFSET*(epsx)$);
\coordinate (Fback) at ($(F) + (epsy) - (epsz) - \OFFSET*(epsx)$);

\coordinate (Etop) at ($(E) + (epsx) + (epsy) + \OFFSET*(epsz)$);
\coordinate (Ftop) at ($(F) - (epsx) + (epsy)  + \OFFSET*(epsz)$);
\coordinate (Gtop) at ($(G) - (epsx) - (epsy)  + \OFFSET*(epsz)$);
\coordinate (Htop) at ($(H) + (epsx) - (epsy)  + \OFFSET*(epsz)$);

  \filldraw[fill=lightgray!80, draw=lightgray!80, very thin] (Abase) -- (Bbase) -- (Cbase) -- (Dbase) -- cycle;

  \draw[{Stealth[harpoon]}-,\FC,dashed] (Abase) -- (Bbase);
  \draw[{Stealth[harpoon]}-,\FC,dashed] (Bbase) -- (Cbase);
  \draw[{Stealth[harpoon]}-,\FC,dashed] (Cbase) -- (Dbase);
  \draw[{Stealth[harpoon]}-,black,very thick] (Dbase) -- (Abase) node[midway,below] {0};
    
    \draw[{Stealth[harpoon]}-,\BC,dashed] (Bleft) -- (Aleft);
    \draw[{Stealth[harpoon]}-,\BC,dashed] (Fleft) -- (Bleft);
    \draw[{Stealth[harpoon]}-,\BC,dashed] (Eleft) -- (Fleft);
    \draw[{Stealth[harpoon]}-,\BC,dashed] (Aleft) -- (Eleft);

    \draw[{Stealth[harpoon]}-,\BC,dashed] (Cback) -- (Bback);
    \draw[{Stealth[harpoon]}-,\BC,dashed] (Bback) -- (Fback);
    \draw[{Stealth[harpoon]}-,\BC,dashed] (Fback) -- (Gback);
    \draw[{Stealth[harpoon]}-,\BC,dashed] (Gback) -- (Cback);

    \draw[{Stealth[harpoon,swap]}-,blue] (Afront) -- (Dfront) node[midway,above] {\color{black} $1$};
    \draw[{Stealth[harpoon,swap]}-,blue] (Dfront) -- (Hfront) node[midway,left] {\color{black} $4$};
    \draw[{Stealth[harpoon,swap]}-,blue] (Hfront) -- (Efront) node[midway,below] {\color{black} $3$};
    \draw[{Stealth[harpoon,swap]}-,blue] (Efront) -- (Afront) node[midway,right] {\color{black} $2$};

    \draw[{Stealth[harpoon,swap]}-,\BC,dashed] (Dright) -- (Cright);
    \draw[{Stealth[harpoon,swap]}-,\BC,dashed] (Cright) -- (Gright);
    \draw[{Stealth[harpoon,swap]}-,\BC,dashed] (Gright) -- (Hright);
    \draw[{Stealth[harpoon,swap]}-,\BC,dashed] (Hright) -- (Dright);

  \draw[{Stealth[harpoon,swap]}-,\BC,dashed] (Htop) -- (Gtop);
  \draw[{Stealth[harpoon,swap]}-,\BC,dashed] (Gtop) -- (Ftop);
  \draw[{Stealth[harpoon,swap]}-,\BC,dashed] (Ftop) -- (Etop);
  \draw[{Stealth[harpoon,swap]}-,blue] (Etop) -- (Htop) node[midway,above right] {\color{black} $5$};

\end{tikzpicture}
}
	\caption[Twenty-four darts of a three-dimensional tensor product domain.]{Twenty-four darts of a three-dimensional tensor product domain are implicitly represented through four two-dimensional representative darts.
	One such dart, on the base, and its corresponding generated darts, are highlighted here using solid lines.}
	\label{fig:TP3d}
\end{figure}
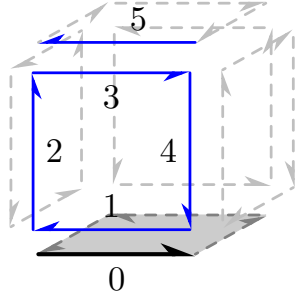

The $\phi$ operations for this tensor product C-map depend on some choices for the dart indexing, including the choice for how to flatten the indexing in the equation above.
For each $\Dim\in\{2,3\}$ we define a template that shows how the $2n$ darts corresponding to each $(\dart_{\mathrm{base}}, \dart_\mathrm{line})$ pair are arranged.
If $\Dim=2$, this corresponds to the four darts in a face, as shown in \cref{fig:TP2d}, and these darts are ordered such that
\begin{equation}
    \phi_1(\dart_\mathrm{tp})=\textrm{flatten}(\dart_\mathrm{base},\dart_\mathrm{line},(a+1)\bmod 4),
\end{equation}
where $(\dart_\mathrm{base},\dart_\mathrm{line},a)=\textrm{unflatten}(\dart_\mathrm{tp})$.
In contrast, for $\Dim=3$, the template includes six darts, four of which correspond to a quadrilateral face which is transverse to the base C-map, and two of which are on opposite faces in the volume, both of which faces are parallel to the base C-map in a parametric sense.
The arrangement of these six darts is shown in \cref{fig:TP3d}.
We can see that for the $\Dim=3$ case the chosen arrangement defines a subset of the $\phi_1$ operations, as well as a subset of the $\phi_2$ operations.
The rest can be defined using $\phi$ operations on $\cmap_\mathrm{base}$.
This is expanded upon in \cref{alg:tp3dphi1,alg:tp3dphi2}.

\begin{figure}
	\centering
\def\FC{gray}
\def\BC{lightgray}

\resizebox{0.25\textwidth}{!}{%
\begin{tikzpicture}

\pgfmathsetmacro{\EPS}{0.1}
\pgfmathsetmacro{\OFFSET}{0.4}

\coordinate (A) at (0,0,0);
\coordinate (B) at (2,0,0);
\coordinate (C) at (2,2,0);
\coordinate (D) at (0,2,0);
\coordinate (E) at (0,0,1);
\coordinate (F) at (1,0,1);
\coordinate (G) at (1,1,1);
\coordinate (H) at (0,1,1);
\coordinate (epsx) at (\EPS,0,0);
\coordinate (epsy) at (0,\EPS,0);
\coordinate (epsz) at (0,0,\EPS);
\coordinate (shift) at (0,1.5,0);

\coordinate (Abase) at ($(A) + (epsx) + (epsy) - \OFFSET*(epsz) $);
\coordinate (Bbase) at ($(B) - (epsx) + (epsy) - \OFFSET*(epsz)$);
\coordinate (Cbase) at ($(C) - (epsx) - (epsy) - \OFFSET*(epsz)$);
\coordinate (Dbase) at ($(D) + (epsx) - (epsy) - \OFFSET*(epsz)$);

%
%
%
%

  \filldraw[fill=lightgray!80, draw=lightgray!80, very thin] (Abase) -- (Bbase) -- (Cbase) -- (Dbase) -- cycle;

  \draw[-{Stealth[harpoon]},black, thick] (Abase) -- (Bbase) node[midway,above] {\color{black}0};
  \draw[-{Stealth[harpoon]},blue] (Bbase) -- (Cbase) node[midway,left] {\color{black}1};
  \draw[-{Stealth[harpoon]},blue] (Cbase) -- (Dbase) node[midway,below] {\color{black}2};
  \draw[-{Stealth[harpoon]},blue] (Dbase) -- (Abase) node[midway,right] {\color{black}3};
    
%
%
%
%
%

\end{tikzpicture}
}
	\caption[Four darts of a two-dimensional tensor product domain.]{Four darts of a two-dimensional tensor product domain are implicitly represented through a one-dimensional representative dart.
	Queries about connectivity information can be passed through a base one-dimensional dart though a templated representation of an associated two-dimensional cell.}
	\label{fig:TP2d}
\end{figure}
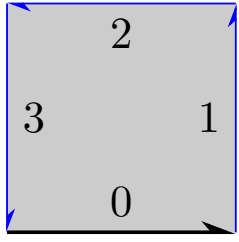

\begin{algorithm}
    \caption{$\phi_1$ ops for tensor product with $\Dim=3$}\label{alg:tp3dphi1}
    \begin{algorithmic} [1]
    \Function{TP3dPhi1}{$\dart$}
    \State $\dart_\mathrm{base},\dart_\mathrm{line},a=\textrm{unflatten}(\dart)$
    \If{$a$ is $5$}
    \State \Return $\textrm{flatten}(\phi_{-1}[\cmap_\mathrm{base}](\dart_\mathrm{base}),\dart_\mathrm{line},a)$
    \ElsIf{$a$ is $0$}
    \State \Return $\textrm{flatten}(\phi_{1}[\cmap_\mathrm{base}](\dart_\mathrm{base}),\dart_\mathrm{line},a)$
    \Else
    \State \Return $\textrm{flatten}(\dart_\mathrm{base},\dart_\mathrm{line},(a\bmod 4) + 1)$
    \EndIf
    \EndFunction
    \end{algorithmic}
\end{algorithm}

\begin{algorithm}
    \caption{$\phi_2$ ops for tensor product with $\Dim=3$}\label{alg:tp3dphi2}
    \begin{algorithmic} [1]
    \Function{TP3dPhi2}{$\dart$}
    \State $\dart_\mathrm{base},\dart_\mathrm{line},a=\textrm{unflatten}(\dart)$
    \State $a'=[1,0,4,5,2,3]\text{.at}(a)$
    \If{$a$ is $2$}
    \State \Return $\textrm{flatten}(\phi_{-1}[\cmap_\mathrm{base}](\dart_\mathrm{base}),\dart_\mathrm{line},a')$
    \ElsIf{$a$ is $4$}
    \State \Return $\textrm{flatten}(\phi_{1}[\cmap_\mathrm{base}](\dart_\mathrm{base}),\dart_\mathrm{line},a')$
    \Else
    \State \Return $\textrm{flatten}(\dart_\mathrm{base},\dart_\mathrm{line},a')$
    \EndIf
    \EndFunction
    \end{algorithmic}
\end{algorithm}

The $\phi_n$ operations move to a new $n$-cell, and as such always involve a $\phi$ operation from one of the C-maps of which $\cmap_\mathrm{tp}$ is composed.
Because these are structured meshes, we have a consistent orientation of every cell with respect to its neighbor, so we can explicitly define the permutation of $a$ that the operation causes.
In \cref{alg:tp2dphi2} we define the $\phi_2$ operations for a 2d $\cmap_\mathrm{tp}$,
\begin{algorithm}
    \caption{$\phi_2$ ops for tensor product with $\Dim=2$}\label{alg:tp2dphi2}
    \begin{algorithmic} [1]
        \Function{TP2dPhi2}{$\dart$}
        \State $\dart_\mathrm{base},\dart_\mathrm{line},a=\textrm{unflatten}(\dart)$
        \State $a'=[2,3,0,1]\text{.at}(a)$ \Comment{Permute $a$}
        \If{$a$ is $0$}
        \State \Return $\textrm{flatten}(\dart_\mathrm{base},\phi_{-1}[\cmap_\mathrm{line}](\dart_\mathrm{line}),a')$
        \ElsIf{$a$ is $1$}
        \State \Return $\textrm{flatten}(\phi_{1}[\cmap_\mathrm{base}](\dart_\mathrm{base}),\dart_\mathrm{line},a')$
        \ElsIf{$a$ is $2$}
        \State \Return $\textrm{flatten}(\dart_\mathrm{base},\phi_{1}[\cmap_\mathrm{line}](\dart_\mathrm{line}),a')$
        \Else
        \State \Return $\textrm{flatten}(\phi_{-1}[\cmap_\mathrm{base}](\dart_\mathrm{base}),\dart_\mathrm{line},a')$
        \EndIf
        \EndFunction
    \end{algorithmic}
\end{algorithm}
and similarly in \cref{alg:tp3dphi3} we define $\phi_3$ operations for a 3d $\cmap_\mathrm{tp}$.
\begin{algorithm}
    \caption{$\phi_3$ ops for tensor product with $\Dim=3$}\label{alg:tp3dphi3}
    \begin{algorithmic} [1]
        \Function{TP3dPhi3}{$\dart$}
        \State $\dart_\mathrm{base},\dart_\mathrm{line},a=\textrm{unflatten}(\dart)$
        \State $a'=[5,1,4,3,2,0]\text{.at}(a)$ \Comment{Permute $a$}
        \If{$a$ is $0$}
        \State \Return $\textrm{flatten}(\dart_\mathrm{base},\phi_{-1}[\cmap_\mathrm{line}](\dart_\mathrm{line}),a')$
        \ElsIf{$a$ is $5$}
        \State \Return $\textrm{flatten}(\dart_\mathrm{base},\phi_{1}[\cmap_\mathrm{line}](\dart_\mathrm{line}),a')$
        \Else
        \State \Return $\textrm{flatten}(\phi_{2}[\cmap_\mathrm{base}](\dart_\mathrm{base}),\dart_\mathrm{line},a')$
        \EndIf
        \EndFunction
    \end{algorithmic}
\end{algorithm}

Note that to this point, the entire definition of the combinatorial map is held in $\Dim$ 1d C-maps---which each hold only a single unsigned integer encoding their size and a boolean indicating periodicity---and a few short static arrays to define permutations on $a$.
This storage is constant as the density of the mesh scales, and is much cheaper than storing lists of size $\numdarts{\cmap_\mathrm{tp}}$ for each $\phi_i,\;i\in\{-1,1,...,\Dim\}$.
This allows us to store the full mesh on each refinement level without concern for memory use.

The algorithms to build up a hierarchical C-map that we will present in the later sections rely on two operations from the set of refinement level C-maps, in addition to those which define a C-map, namely, given a dart from the hierarchical dart index space, the ability to find an ancestor dart on a lower level if one exists, and the ability to iterate the descendant darts on a higher level.
We will combine these into one routine that may return several descendant darts, one dart if an ancestor exists, or zero darts if no ancestor exists, and call it $\textproc{DartLineage}(\dart,\ell)$, where $\ell$ is the level on which relative darts are requested.
For the case of the tensor-product combinatorial maps, this can be implemented by unflattening the given dart to $\Dim$ darts and multiplying by, dividing by, or taking the modulus with respect to the ratio of numbers of darts in the two levels in the appropriate parametric dimensions.
The modulus operator answers the question of if an ancestor exists, division finds the darts that can be flattened to create the ancestor, and multiplying gives the starting index of the darts that can be flattened to create the descendant darts.

\subsubsection{Multi-patch}

In the context of B-splines and hierarchical B-splines, multi-patch typically means a collection of tensor-product regions (called patches) connected to one another along the entire portion of the boundary where one of the parametric coordinates is a constant $0$ or $1$.
The basis across that connected interface is typically only $C^0$ smooth at a maximum.
For the C-map, the question of the basis is unimportant, and we call a C-map multi-patch if it is a representation of an unstructured or semi-structured quadrilateral or hexahedral mesh by decomposition into tensor-product regions.
These regions, as in the case of B-splines, should be connected along the entire portion of the boundary over which one of the parametric coordinates is constant, or in other words, one of the four sides in 2d, or one of the six sides in 3d.
Arbitrarily smooth splines (where formulations exist to build them) can be represented over these meshes as long as they are built over conforming hexahedral or quadrilateral B\'ezier meshes.

We form a multi-patch C-map $\cmap_\mathrm{mp}$ from a collection of $\NumPatches$ tensor-product C-maps $\{\cmap_{\mathrm{tp},i}\}_{i=0}^{\NumPatches-1}$.
The darts in $\cmap_\mathrm{mp}$ have a one-to-one correspondence with the darts in $\{\cmap_{\mathrm{tp},i}\}_{i=0}^{\NumPatches-1}$, and we flatten those darts into a common index space as
\begin{equation}\label{eq:mp_dart_from_tp_dart}
\dart_\mathrm{mp}=\dart_\mathrm{tp} + \sum_{i=0}^{q-1}\numdarts{\cmap_{\mathrm{tp},i}},
\end{equation}
where $q$ is the number of the patch from which $\dart_\mathrm{tp}$ comes.
We can form the inverse mapping by storing at construction the sorted array of partial sums $\left[\sum_{i=0}^{j}\numdarts{\cmap_{\mathrm{tp},i}}\right]_{j=0}^{\NumPatches-1}$, performing an $\mathcal{O}(\log n_p)$ search for the last entry  of this sorted list that is less than $\dart_\mathrm{mp}$, and subtracting that from $\dart_\mathrm{mp}$.
Since $\NumPatches\ll\numcells{\cmap_\mathrm{mp}}$ for the semi-structured meshes necessary for effective use of splines, this is essentially a constant-time operation with respect to mesh size.

\begin{figure}
    \centering
\def\FC{black}
\def\BC{lightgray!40}

\begin{tikzpicture}[
  thick, line cap=round, line join=round
]
\pgfmathsetmacro{\EPS}{0.06}
\pgfmathsetmacro{\OFFSET}{0.4}

\begin{scope}[
  x={(1.3cm,0.6cm)},
  y={(1.5cm,-0.5cm)},
  z={(0cm,1.8cm)}
]

\coordinate (epsx) at (\EPS,0,0);
\coordinate (epsz) at (0,0,\EPS);

\draw[-,\FC] (0,0,1)--(0,1,1)--(1,1,1)--(1,0,1)--(0,0,1)--(0,0,0)--(0,1,0)--(1,1,0)--(1,1,1) (0,1,1)--(0,1,0);
\draw[-,thin,\FC] (0,0.5,0)--(0,0.5,1)--(1,0.5,1);
\draw[-,thin,\FC] (0,0,0.5)--(0,1,0.5)--(1,1,0.5);
\draw[-,thin,\FC] (0.5,0,1)--(0.5,1,1)--(0.5,1,0);

\draw[{Stealth[harpoon,swap]}-,\FC] ($(0.5,1,0.5)+(epsx)+(epsz)$) -- ($(1,1,0.5)-(epsx)+(epsz)$);

\end{scope}

\begin{scope}[
  x={(1.5cm,0.5cm)},
  y={(1.3cm,-0.6cm)},
  z={(0cm,1.8cm)},
  shift={(5cm,0.1cm)}
]

\coordinate (epsy) at (0,\EPS,0);
\coordinate (epsz) at (0,0,\EPS);

\draw[-,\FC] (0,0,1)--(0,1,1)--(1,1,1)--(1,0,1)--(0,0,1)--(0,0,0)--(0,1,0)--(1,1,0)--(1,1,1) (0,1,1)--(0,1,0);
\draw[-,thin,\FC] (0,0.5,0)--(0,0.5,1)--(1,0.5,1);
\draw[-,thin,\FC] (0,0,0.5)--(0,1,0.5)--(1,1,0.5);
\draw[-,thin,\FC] (0.5,0,1)--(0.5,1,1)--(0.5,1,0);

\draw[{Stealth[harpoon,swap]}-,\FC] ($(0,0,0.5)+(epsy)+(epsz)$) -- ($(0,0.5,0.5)-(epsy)+(epsz)$);

\end{scope}

\begin{scope}[
  x={(1.8cm,0.2cm)},
  y={(0,-1.8cm)},
  shift={(1cm,-1.5cm)}
]
  \coordinate (epsx) at (\EPS,0);
  \coordinate (epsy) at (0,\EPS);

  \draw[-,\FC] (0,0) -- (1,0) -- (1,1) -- (0,1) -- (0,0);
  \draw[-,\FC,thin](0.5,0) -- (0.5,1) (0,0.5)--(1,0.5);

  \draw[{Stealth[harpoon,swap]}-,\FC] ($(0.5,0.5)+(epsx)-(epsy)$) -- ($(1,0.5)-(epsx)-(epsy)$);
\end{scope}

\begin{scope}[
  x={(1.8cm,-0.2cm)},
  y={(0cm,-1.8cm)},
  shift={(5cm,-1.3cm)}
]
  \coordinate (epsx) at (\EPS,0);
  \coordinate (epsy) at (0,\EPS);

  \draw[-,\FC] (0,0) -- (1,0) -- (1,1) -- (0,1) -- (0,0);
  \draw[-,\FC,thin](0.5,0) -- (0.5,1) (0,0.5)--(1,0.5);

  \draw[{Stealth[harpoon,swap]}-,\FC] ($(0,0.5)+(epsx)-(epsy)$) -- ($(0.5,0.5)-(epsx)-(epsy)$);
\end{scope}

\draw[-Stealth] (3cm,1.4cm) to[bend left=30] node[above] {$\phi_3$} (4.8cm,1.4cm);
\draw[-Stealth] (0.5cm,-0.5cm) to[bend right=30] node[left] {drop} (0.8cm,-2cm);
\draw[-Stealth] (3.2cm,-2.8cm) to[bend right=15] node[below] {permute} (4.6cm,-2.8cm);
\draw[-Stealth] (7.3cm,-2cm) to[bend right=30] node[right] {raise} (7.3cm,-0.5cm);

\end{tikzpicture}
    \caption{A $\phi_3[\cmap_\mathrm{mp}](\cdot)$ operation between two 3d patches shown in an exploded view. The $\cmap_\mathrm{mp}$ calculates $\phi_\Dim$ operations between two patches like this by dropping a coordinate from the boundary containing the dart being operated on to find the 2d representation of the dart, permuting the dart to the 2d dart representing the result of the $\phi_\Dim$ operation, and then raising the dart to the final 3d representation.}
    \label{fig:multipatch_phi3}
\end{figure}
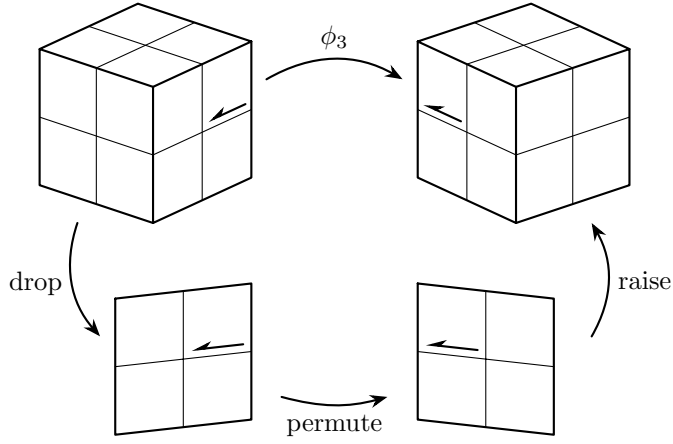

All of the $\phi_i[\cmap_\mathrm{mp}](\cdot)$ operations come from the underlying $\phi_i[\cmap_{\mathrm{tp},p}](\cdot)$ operations, except for some $\phi_n$ operations for darts that are on the boundary in their respective $\cmap_{\mathrm{tp},i}$ but are not on the boundary in $\cmap_\mathrm{mp}$.
For this purpose, we also provide a set of inter-patch connections for the construction of $\cmap_\mathrm{mp}$.
As illustrated in \cref{fig:multipatch_phi3}, we encode $\phi_\Dim$ operations between neighboring patches by (1) dropping one coordinate from the dart to an effective $(\Dim-1)$-dimensional C-map representing the side of the patch, (2) permuting the dart to its counterpart in the side of the adjacent patch, and (3) adding one coordinate to bring it back to the $n$-dimensional patch on the other side of the interface.
The dropping and adding of coordinates is independent of the specific connection, so for each connection we need only to store the side and patch indices of each patch that the connection is between and the permutation between the two sides.
Permutations are performed on the $\Dim-1$ darts obtained from unflattening the dart in the $(\Dim-1)$-dimensional side.
There are only four possible permutations in a 2d side of a 3d patch, and only one permutation in the 1d side of a 2d patch, and the permutation operations can be stored as a simple enumerated type distinguishing between the five types of permutations.
For ease of lookup, we store each connection twice, keyed on each of the two sides involved.

\subsection{Hierarchical}\label{sec:hierarchical_algorithm}

\begin{figure*}[p]
    \centering
\def\FC{black}
\def\BC{gray}
\def\UC{red!75!black}

\resizebox{0.98\textwidth}{!}{%
\begin{tikzpicture}[
  x={(4.5cm,0cm)},        
  y={(0cm,4.5cm)},        
  very thick, line cap=round, line join=round
]

\pgfmathsetmacro{\EPS}{0.02}


\coordinate (A) at (0,0);
\coordinate (B) at (1,0);
\coordinate (C) at (1,1);
\coordinate (D) at (0,1);
\coordinate (E) at (0,2);
\coordinate (F) at (1,2);
\coordinate (epsx) at (\EPS,0);
\coordinate (epsy) at (0,\EPS);
\coordinate (shift) at (1.2,0);

\coordinate (Abot) at ($(A) + (epsx) + (epsy)$);
\coordinate (Bbot) at ($(B) - (epsx) + (epsy)$);
\coordinate (Cbot) at ($(C) - (epsx) - (epsy)$);
\coordinate (Dbot) at ($(D) + (epsx) - (epsy)$);

\coordinate (Dtop) at ($(D) + (epsx) + (epsy)$);
\coordinate (Ctop) at ($(C) - (epsx) + (epsy)$);
\coordinate (Ftop) at ($(F) - (epsx) - (epsy)$);
\coordinate (Etop) at ($(E) + (epsx) - (epsy)$);

\draw[-{Stealth[harpoon]},\FC] (Abot) -- node[font=\small,above, text=black] {0} (Bbot);
\draw[-{Stealth[harpoon]},\FC] (Bbot) -- node[font=\small,left, text=black]  {1} (Cbot);
\draw[-{Stealth[harpoon]},\FC] (Cbot) -- node[font=\small,below, text=black] {2} (Dbot);
\draw[-{Stealth[harpoon]},\FC] (Dbot) -- node[font=\small,right, text=black] {3} (Abot);

\draw[-{Stealth[harpoon]},\UC] (Dtop) -- node[font=\small,above, text=black] {4} (Ctop);
\draw[-{Stealth[harpoon]},\UC] (Ctop) -- node[font=\small,left, text=black]  {5} (Ftop);
\draw[-{Stealth[harpoon]},\UC] (Ftop) -- node[font=\small,below, text=black] {6} (Etop);
\draw[-{Stealth[harpoon]},\UC] (Etop) -- node[font=\small,right, text=black] {7} (Dtop);


\coordinate (G) at (0.5,0);
\coordinate (J) at (0,0.5);
\coordinate (N) at (0.5,0.5);


\foreach \i in {0,1} {
  \foreach \j in {0,1,...,3} {

    \ifnum\j<2
      \ifnum\i=1
        \ifnum\j=1
          \def\linecolor{\FC}
        \else
          \def\linecolor{\UC}
        \fi
      \else
        \def\linecolor{\UC}
      \fi
    \else
      \def\linecolor{\FC}
    \fi

    \coordinate (A11) at ($(A) + (epsx) + (epsy) + (shift) + \i*(0.5,0) + \j*(0,0.5)$);
    \coordinate (G11) at ($(G) - (epsx) + (epsy) + (shift) + \i*(0.5,0) + \j*(0,0.5)$);
    \coordinate (N11) at ($(N) - (epsx) - (epsy) + (shift) + \i*(0.5,0) + \j*(0,0.5)$);
    \coordinate (J11) at ($(J) + (epsx) - (epsy) + (shift) + \i*(0.5,0) + \j*(0,0.5)$);

    \pgfmathtruncatemacro{\dartbottom}{4*(\i + 2*\j)}
    \pgfmathtruncatemacro{\dartright}{\dartbottom+1}
    \pgfmathtruncatemacro{\darttop}{\dartbottom+2}
    \pgfmathtruncatemacro{\dartleft}{\dartbottom+3}

    \draw[-{Stealth[harpoon]},\linecolor] (A11) -- node[font=\small,above, text=black] {\dartbottom} (G11);
    \draw[-{Stealth[harpoon]},\linecolor] (G11) -- node[font=\small,left, text=black]  {\dartright} (N11);
    \draw[-{Stealth[harpoon]},\linecolor] (N11) -- node[font=\small,below, text=black] {\darttop} (J11);
    \draw[-{Stealth[harpoon]},\linecolor] (J11) -- node[font=\small,right, text=black] {\dartleft} (A11);
  }
}


\foreach \i in {0,1,...,5} {
  \foreach \j in {0,1,...,7} {

    \ifnum\i>2
      \ifnum\j>1
        \ifnum\j<4
          \def\linecolor{\UC}
        \else
          \def\linecolor{\FC}
        \fi
      \else
        \def\linecolor{\FC}
      \fi
    \else
      \def\linecolor{\FC}
    \fi

    \coordinate (W) at ($(A) + (epsx) + (epsy) + 2*(shift) + \i*(0.1666,0) + \j*(0,0.25)$);
    \coordinate (X) at ($(0.1666,0) - (epsx) + (epsy) + 2*(shift) + \i*(0.1666,0) + \j*(0,0.25)$);
    \coordinate (Y) at ($(0.1666,0.25) - (epsx) - (epsy) + 2*(shift) + \i*(0.1666,0) + \j*(0,0.25)$);
    \coordinate (Z) at ($(0,0.25) + (epsx) - (epsy) + 2*(shift) + \i*(0.1666,0) + \j*(0,0.25)$);

    \pgfmathtruncatemacro{\dartbottom}{4*(\i + 6*\j)}

    \draw[-{Stealth[harpoon]},\linecolor] (W) -- node[font=\small,above, text=black] {\dartbottom} (X);
    \draw[-{Stealth[harpoon]},\linecolor] (X) -- (Y);
    \draw[-{Stealth[harpoon]},\linecolor] (Y) -- (Z);
    \draw[-{Stealth[harpoon]},\linecolor] (Z) -- (W);
  }
}

\node[above] at (0.5,2) {$\cmap_0$};
\node[above] at ($(0.5,2) + (shift)$) {$\cmap_1$};
\node[above] at ($(0.5,2) + 2*(shift)$) {$\cmap_2$};
\node[above] at ($(0.5,-0.5)+ (shift)$) {$\cmap_h$};


\coordinate (shifth) at ($(shift) - (0,2.5)$);
\coordinate (Dtop_shifted) at ($(D) + (epsx) + (epsy) + (shifth)$);
\coordinate (Ctop_shifted) at ($(C) - (epsx) + (epsy) + (shifth)$);
\coordinate (Ftop_shifted) at ($(F) - (epsx) - (epsy) + (shifth)$);
\coordinate (Etop_shifted) at ($(E) + (epsx) - (epsy) + (shifth)$);

\draw[-{Stealth[harpoon]},\FC] (Ctop_shifted) -- node[font=\small,left] {5} (Ftop_shifted);
\draw[-{Stealth[harpoon]},\FC] (Ftop_shifted) -- node[font=\small,below] {6} (Etop_shifted);
\draw[-{Stealth[harpoon]},\FC] (Etop_shifted) -- node[font=\small,right] {7} (Dtop_shifted);

\coordinate (Ahier) at ($(A) + (epsx) + (epsy) + (shifth)$);
\coordinate (Ghier) at ($(G) - (epsx) + (epsy) + (shifth)$);
\coordinate (Nhier) at ($(N) - (epsx) - (epsy) + (shifth)$);
\coordinate (Jhier) at ($(J) + (epsx) - (epsy) + (shifth)$);

\draw[-{Stealth[harpoon]},\FC] (Ahier) -- node[font=\small,above] {8} (Ghier);
\draw[-{Stealth[harpoon]},\FC] (Ghier) -- node[font=\small,left]  {9} (Nhier);
\draw[-{Stealth[harpoon]},\FC] (Nhier) -- node[font=\small,below] {10} (Jhier);
\draw[-{Stealth[harpoon]},\FC] (Jhier) -- node[font=\small,right] {11} (Ahier);

\coordinate (Ashifted) at ($(Ahier) + (0.5,0)$);
\coordinate (Gshifted) at ($(Ghier) + (0.5,0)$);
\coordinate (Nshifted) at ($(Nhier) + (0.5,0)$);
\coordinate (Jshifted) at ($(Jhier) + (0.5,0)$);

\draw[-{Stealth[harpoon]},\FC] (Ashifted) -- node[font=\small,above] {12} (Gshifted);
\draw[-{Stealth[harpoon]},\FC] (Gshifted) -- node[font=\small,left]  {13} (Nshifted);
\draw[-{Stealth[harpoon]},\FC] (Jshifted) -- node[font=\small,right] {15} (Ashifted);

\coordinate (Ashifted) at ($(Ahier) + (0,0.5)$);
\coordinate (Gshifted) at ($(Ghier) + (0,0.5)$);
\coordinate (Nshifted) at ($(Nhier) + (0,0.5)$);
\coordinate (Jshifted) at ($(Jhier) + (0,0.5)$);

\draw[-{Stealth[harpoon]},\FC] (Ashifted) -- node[font=\small,above] {16} (Gshifted);
\draw[-{Stealth[harpoon]},\FC] (Nshifted) -- node[font=\small,below] {18} (Jshifted);
\draw[-{Stealth[harpoon]},\FC] (Jshifted) -- node[font=\small,right] {19} (Ashifted);

\coordinate (Ashifted) at ($(Ahier) + (0,1)$);
\coordinate (Gshifted) at ($(Ghier) + (0,1)$);

\draw[-{Stealth[harpoon]},\FC] (Ashifted) -- node[font=\small,above] {24} (Gshifted);

\coordinate (W) at ($(A) + (epsx) + (epsy) + (shifth)$);
\coordinate (X) at ($(0.1666,0) - (epsx) + (epsy) + (shifth)$);
\coordinate (Y) at ($(0.1666,0.25) - (epsx) - (epsy) + (shifth)$);
\coordinate (Z) at ($(0,0.25) + (epsx) - (epsy) + (shifth)$);

\foreach \i in {3,4,5} {
  \foreach \j in {2,3} {
    \coordinate (W) at ($(A) + (epsx) + (epsy) + (shifth) + \i*(0.1666,0) + \j*(0,0.25)$);
    \coordinate (X) at ($(0.1666,0) - (epsx) + (epsy) + (shifth) + \i*(0.1666,0) + \j*(0,0.25)$);
    \coordinate (Y) at ($(0.1666,0.25) - (epsx) - (epsy) + (shifth) + \i*(0.1666,0) + \j*(0,0.25)$);
    \coordinate (Z) at ($(0,0.25) + (epsx) - (epsy) + (shifth) + \i*(0.1666,0) + \j*(0,0.25)$);

    \pgfmathtruncatemacro{\dartbottom}{4*(\i + 6*\j)+40}
    \pgfmathtruncatemacro{\dartright}{\dartbottom+1}
    \pgfmathtruncatemacro{\darttop}{\dartbottom+2}
    \pgfmathtruncatemacro{\dartleft}{\dartbottom+3}
    \draw[-{Stealth[harpoon]},\FC] (W) -- node[font=\small,above] {\dartbottom} (X);
    \draw[-{Stealth[harpoon]},\FC] (X) -- node[font=\small,left]  {} (Y);
    \draw[-{Stealth[harpoon]},\FC] (Y) -- node[font=\small,below] {} (Z);
    \draw[-{Stealth[harpoon]},\FC] (Z) -- node[font=\small,right] {} (W);
  }
}

\foreach \j in {2,3} {
  \coordinate (X) at ($(0.1666,0) - (epsx) + (epsy) + (shifth) + 2*(0.1666,0) + \j*(0,0.25)$);
  \coordinate (Y) at ($(0.1666,0.25) - (epsx) - (epsy) + (shifth) + 2*(0.1666,0) + \j*(0,0.25)$);

  \pgfmathtruncatemacro{\dart}{97+24*(\j-2)}
  \draw[-{Stealth[harpoon]},\FC] (X) -- node[font=\small,left]{\dart} (Y);
}

\foreach \i in {3,4,5}{
  \coordinate (W) at ($(A) + (epsx) + (epsy) + (shifth) + \i*(0.1666,0) + 4*(0,0.25)$);
  \coordinate (X) at ($(0.1666,0) - (epsx) + (epsy) + (shifth) + \i*(0.1666,0) + 4*(0,0.25)$);
  \coordinate (Y) at ($(0.1666,0.25) - (epsx) - (epsy) + (shifth) + \i*(0.1666,0) + 1*(0,0.25)$);
  \coordinate (Z) at ($(0,0.25) + (epsx) - (epsy) + (shifth) + \i*(0.1666,0) + 1*(0,0.25)$);

  \pgfmathtruncatemacro{\darttop}{148+4*(\i-3)}
  \pgfmathtruncatemacro{\dartbot}{78+4*(\i-3)}
  \draw[-{Stealth[harpoon]},\FC] (W) -- node[font=\small,above] {\darttop} (X);
  \draw[-{Stealth[harpoon]},\FC] (Y) -- node[font=\small,below] {\dartbot} (Z);
}

\end{tikzpicture}
}
    \caption{Example construction of $\cmap_h$. The global refinement levels $\cmap_0$, $\cmap_1$, and $\cmap_2$ are also shown. Darts are labelled according to the indexing defined in \cref{sec:data_structures}, save that in $\cmap_2$ (and in the portions of $\cmap_h$ whose darts come from $\cmap_2$) some dart labels are omitted for lack of space where they can be inferred from the surrounding labels. Darts marked in red are those which are darts of leaf elements, or in other words, those included in $\UnrefinedLeafDarts$.}
    \label{fig:hierCmap}
\end{figure*}
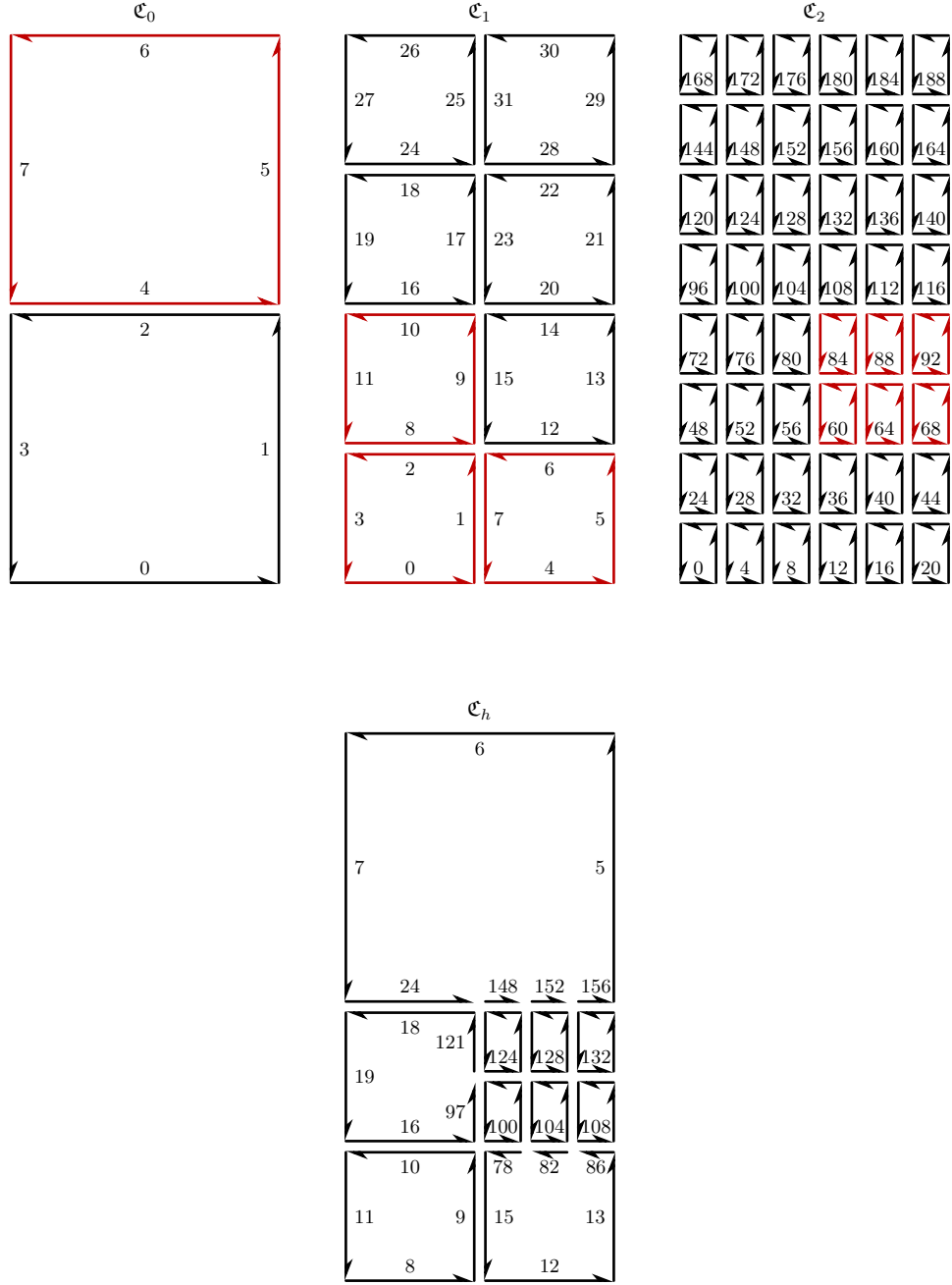

A hierarchical C-map $\cmap_h$ is the C-map representing the B\'ezier mesh of hierarchical B-splines or truncated hierarchical B-splines.
It can also represent the B\'ezier mesh of a hierarchical refinement of other splines that can be represented by a conforming quadrilateral or hexahedral B\'ezier mesh on each level.
An example where it falls down for nonconforming B\'ezier meshes is given in appendix \ref{sec:tjunctionfail}.
The combinatorial map for a hierarchical B-spline basis (whether truncated or not) contains a subset of darts from $\NumRefinementLevels$ tensor-product or multi-patch C-maps $\{\cmap_{i}\}_{i=0}^{\NumRefinementLevels-1}$.
\Cref{fig:hierCmap} demonstrates this for a $\cmap_h$ with $\NumRefinementLevels=3$, where $\cmap_h$ contains three darts from $\cmap_0$, eleven darts from $\cmap_1$, and 32 darts from $\cmap_2$.
We allow the darts in the hierarchical C-map to be non-contiguous: while $\numdarts{\cmap_h}\leq\sum_{i=0}^{\NumRefinementLevels-1}\numdarts{\cmap_{i}}$ (with the equality holding only when $\NumRefinementLevels<2$), dart indices come from the range $\left[0,\sum_{i=0}^{\NumRefinementLevels-1}\numdarts{\cmap_{i}}-1\right]$.
We find no downsides to the non-contiguous dart indexing, save perhaps an earlier need to represent dart indices as 64-bit unsigned integers rather than 32-bit.
This index space gives a natural mapping between the level and index of a dart in one of the refinement levels, and its potential analog in the hierarchical C-map:
\begin{equation}\label{eq:hier_dart_from_level_dart}
\dart_h=\operatorname{hierDart}(\mathsf{d},\ell):=\dart + \sum_{i=0}^{\ell-1}\numdarts{\cmap_{i}},
\end{equation}
where $\ell$ is the refinement level of the dart $\mathsf{d}$.
Taking as an example the darts in \cref{fig:hierCmap}, dart 5 in $\cmap_0$ maps to dart 5 in $\cmap_h$, while dart 16 from $\cmap_1$ maps to dart 24 in $\cmap_h$, and dart 108 in $\cmap_2$ maps to dart 148 in $\cmap_h$.
As in the multipatch construction, we store at construction the sorted list of partial sums $\left[\sum_{i=0}^{j}\numdarts{\cmap_{i}}\right]_{j=0}^{\NumRefinementLevels-1}$ to enable the inverse mapping by performing an $\mathcal{O}(\log n_\ell)$ search for the last entry of the sorted list that is less than $\dart_h$, and subtracting that from $\dart_h$.
Since $n_\ell\ll\numcells{\cmap_h}$, this is once again essentially a constant-time operation with respect to mesh size.

To indicate which of the darts in this larger index space are leaf darts, meaning actually in the hierarchical C-map, we store a bitmap $\LeafDarts$ that indicates true or false on each index in the range $0,\ldots,\sum_{i=0}^{\NumRefinementLevels-1}\numdarts{\cmap_{i}}-1$.
Because only a small fraction of the darts in most refinement levels are active, this bitmap is a prime candidate for compression through techniques such as Roaring Bitmaps \cite{chambi_better_2016}.
A hierarchical basis is specified through the domains $\Omega_i$, typically using $\NumRefinementLevels$ sets of refinement-level leaf elements $\{\operatorname{leaf}_i\}_{i=0}^{\NumRefinementLevels-1}$, which are the elements from each $\cmap_i$ that are included in the final mesh of $\cmap_h$.
The domains $\Omega_i$ can then be defined as
\begin{equation*}
    \Omega_i=\bigcup_{j=i}^{\NumRefinementLevels-1}\bigcup_{\mathsf{c}\in\operatorname{leaf}_j}\overline{F(\mathsf{c})}.
\end{equation*}
We compute from the leaf elements another bitmap $\UnrefinedLeafDarts$ which stores if each dart is contained in a leaf element in its refinement level.
More precisely, this means that
\begin{equation*}
    \UnrefinedLeafDarts = \bigcup_{i=0}^{\NumRefinementLevels-1}\{\operatorname{hierDart}(\mathsf{d},i)\,:\,\mathsf{d}\in\operatorname{dartsOf}(\mathsf{c}),\mathsf{c}\in\operatorname{leaf}_i\}.
\end{equation*}
In the example of \cref{fig:hierCmap}, the bitmap $\LeafDarts$ stores all the darts shown in $\cmap_h$, but $\UnrefinedLeafDarts$ would instead store the mapping to the greater index space of darts 4 through 7 from $\cmap_0$, 0 through 11 from $\cmap_1$, and 60 through 71 and 84 through 95 from $\cmap_2$, shown in red in the figure.
This bitmap is also a great candidate for compression.

We also define three other helper functions needed based on \textproc{DartLineage} and $\LeafDarts$:
\begin{equation*}
\textproc{AncestorDarts}(\mathsf{d})=\bigcup_{i=0}^{\ell-1}\textproc{DartLineage}(\mathsf{d},i)
\end{equation*}
and
\begin{equation*}
\textproc{LeafDescendants}(\mathsf{d})=
\left(\bigcup_{i=\ell}^{n_\ell-1}\textproc{DartLineage}(\mathsf{d},i)\right)\cap\LeafDarts,
\end{equation*}
where $\ell$ is the refinement level to which $\mathsf{d}$ belongs, and
\begin{equation*}
    \textproc{HasLeafAncestor}(\mathsf{d})=\textproc{AncestorDarts}(\mathsf{d})\cap\LeafDarts\neq\emptyset.
\end{equation*}

\Cref{alg:hierleafdarts} details the process of determining programmatically which darts belong in $\LeafDarts$ and $\UnrefinedLeafDarts$.
It iterates through the leaf elements, adding the darts of those elements, and in the case of $\LeafDarts$ some darts in neighboring cells, and then at the end removes any darts from $\LeafDarts$ whose descendants were added.

\begin{algorithm}
    \caption{Calculation of leaf darts for hierarchical cmaps.}\label{alg:hierleafdarts}
    \begin{algorithmic} [1]
        \Function{LeafDarts}{$\{\cmap_{i}\}_{i=0}^{\NumRefinementLevels-1}$,$\{\operatorname{leaf}_i\}_{i=0}^{\NumRefinementLevels-1}$}
        \State $\NonLeafDarts \gets \{\}$ \Comment{Darts to later mark as non-leaf}
        \State $\UnrefinedLeafDarts \gets \{\}$ \Comment{Unrefined darts in the leaf elements}
        \State $\LeafDarts \gets \{\}$ \Comment{Leaf darts}
        \Statex 
        \Procedure{MarkAsLeaf}{$\dart$,$\ell$} \textbf{using} $\LeafDarts,\NonLeafDarts$
            \State $\LeafDarts\gets\LeafDarts\cup\{\operatorname{hierDart}(\dart,\ell)\}$
            \State $\NonLeafDarts\gets\NonLeafDarts\cup\textproc{ancestorDarts}(\dart)$
        \EndProcedure
        \Statex
        \For{$\ell=0,\ldots,\NumRefinementLevels-1$}
            \For{$\mathsf{d}\in\{\operatorname{dartsOf}(\mathsf{c}):\mathsf{c}\in\operatorname{leaf}_\ell\}$}
                \State $\UnrefinedLeafDarts\gets\UnrefinedLeafDarts\cup\{\operatorname{hierDart}(\dart, \ell)\}$
                \State $\textproc{MarkAsLeaf}(\dart)$
                \If{$\phi_{n}[\cmap_{\mathrm{tp},\ell}](\dart)$ exists}
                    \State $\textproc{MarkAsLeaf}(\phi_{n}[\cmap_{\mathrm{tp},\ell}](\dart))$
                \EndIf
                \For{$\otherdart\in\operatorname{dartsOf}(\operatorname{edge}(\dart))$}
                    \If{$\textproc{hasLeafAncestor}(\otherdart)$}
                        \State $\textproc{MarkAsLeaf}(\otherdart)$
                    \EndIf
                \EndFor
            \EndFor
        \EndFor
        \State $\LeafDarts\gets\LeafDarts\setminus\NonLeafDarts$
        \State \Return $\LeafDarts$, $\UnrefinedLeafDarts$
        \EndFunction
    \end{algorithmic}
\end{algorithm}

Once $\LeafDarts$ is computed, the only missing portion of $\cmap_h$ is some $\phi_1$ and $\phi_{-1}$ operations.
Many of the $\phi_1$ and $\phi_{-1}$ operations and all other $\phi$ operations are simply the composition of \cref{eq:hier_dart_from_level_dart} and its inverse with operations in an underlying refinement level.
We choose a bidirectional map (or bimap) to store these, as keying on the left gives the $\phi_1$ operations, and keying on the right gives the $\phi_{-1}$ operations.
These are computed in \cref{alg:hierphi1}.

\begin{algorithm}
    \caption{Calculation of $\phi_1$ and $\phi_{-1}$ ops for hierarchical cmaps.}\label{alg:hierphi1}
    \begin{algorithmic} [1]
        \Function{HierPhi1}{$\{\cmap_{i}\}_{i=0}^{\NumRefinementLevels-1}$,$\{\operatorname{leaf}_i\}_{i=0}^{\NumRefinementLevels-1}$}
            \State $\mathbf{M}\gets\{\}$\Comment{The bimap storing $\phi_1$ and $\phi_{-1}$ ops.}
            \Statex\Comment{Adds a sequence of $\phi_1$ ops on a face to $\mathbf{M}$}
            \Procedure{phiChain}{$\mathsf{f}$} \textbf{using} $\mathbf{M}$
                \State $\dart\gets \mathsf{f}.\mathrm{dart}()$
                \State $m\gets[]$
                \Repeat
                    \State $m\gets m + \textproc{LeafDescendants}(\dart)$
                    \State $\dart\gets\phi_1(\dart)$
                \Until{ $\dart = \mathsf{f}.\mathrm{dart}()$ }
                \State $\mathbf{M}\gets\mathbf{M}\cup\{m[|m|-1]\leftrightarrow m[0]\}$
                \For{$i=0,\ldots,|m|-2$}
                \State $\mathbf{M}\gets\mathbf{M}\cup\{m[i]\leftrightarrow m[i+1]\}$
                \EndFor
            \EndProcedure
            \Statex
            \For{$\ell=\NumRefinementLevels-2,\ldots,0$}
                \For{$\mathsf{f}\in\{\operatorname{Adj}^2(\mathsf{c}):\mathsf{c}\in\operatorname{leaf}_\ell\}$}
                \If{$\operatorname{hierDart}(\operatorname{dartsOf}(\mathsf{f}),\ell)\subset\LeafDarts$}
                    \State \textbf{continue}
                \EndIf
                \State $\textproc{phiChain}(\mathsf{f})$
                \If{$\phi_{3}[\cmap_{\mathrm{tp},\ell}](\mathsf{f}.\mathrm{dart}())$ exists}
                    \State $\textproc{phiChain}(\operatorname{face}(\phi_{3}[\cmap_{\mathrm{tp},\ell}](\mathsf{f}.\mathrm{dart}())))$
                    \EndIf
                \EndFor
            \EndFor
            \State \Return $\mathbf{M}$
        \EndFunction
    \end{algorithmic}
\end{algorithm}

\subsubsection{Memory Usage}

Given the data structures from the previous section, we can summarize here the memory needed to define the full hierarchical combinatorial map.
An $\Dim$-dimensional hierarchical combinatorial map with $n_\ell$ levels and $n_p$ patches defined as above stores at most $\Dim\times n_p \times n_\ell$ one-dimensional combinatorial maps, which each are represented by an unsigned integer number of elements and a boolean to indicate periodicity, as well as a small amount of data shared between all the maps representing necessary dart permutations for $\phi$ operations.
In addition, if $n_p>1$, there are $n_p\times n_\ell$ dart offsets stored to map patch darts into the multipatch dart index space, and interpatch connections stored as around $10 n_c$ integers, where $n_c\approx n_p$ is the number of interpatch connections.
Finally, the hierarchical combinatorial map stores $n_\ell$ additional dart offsets to map refinement level darts into the hierarchical dart space, as well as the bitmaps $\LeafDarts$ and $\UnrefinedLeafDarts$ and the bidirectional map $\mathbf{M}$.
The bitmaps and the bidirectional map typically dwarf all other memory usage; each bitmap is theoretically capped at $\sum_{i=0}^{\NumRefinementLevels-1}\numdarts{\cmap_{i}}$ bits but with roaring bitmaps will use much less memory than that.

\subsubsection{Example}

Here we show an example hierarchical combinatorial map, and walk through \cref{alg:hierleafdarts,alg:hierphi1} on the example to explain them.
Our example here is a two-dimensional tensor product hierarchical combinatorial map with $\NumRefinementLevels=3$ and $\numcells{\cmap_h}=13$, shown in \cref{fig:hierCmap}.
We use a refinement ratio of 2 between $\cmap_0$ and $\cmap_1$ in both parametric dimensions, and a refinement level of $(3,2)$ between $\cmap_1$ and $\cmap_2$.
Notice also that there are three darts in $\cmap_h$ from $\cmap_0$, eleven from $\cmap_1$, and 32 from $\cmap_2$.

\Cref{alg:hierleafdarts} decides which of the darts from each level are present in $\cmap_h$, given the list of elements from each level that are present in $\cmap_h$.
We can walk through the algorithm as follows:
We start with $\ell=0$, and iterate the darts of $\operatorname{face}(4)$, the upper face in $\cmap_0$ that can be represented by dart 4.
For each dart of the face, we mark it as an unrefined dart as well as a leaf dart.
There are no ancestors of these darts, so we do not add anything to $\NonLeafDarts$ from this level.
We also add one additional dart, dart 2 from $\cmap_0$ because of line 14 in the algorithm.
Lines 16-19 do not add anything additional in 2d.
Then we iterate to $\ell=1$, where we have the set of leaf elements $\operatorname{leaf}_1=\{\operatorname{face}(0),\operatorname{face}(4),\operatorname{face}(8)\}$.
We add darts 0 through 11 from $\cmap_1$ to $\UnrefinedLeafDarts$, darts 0 through 12 and 15 and 16 from $\cmap_1$ to $\LeafDarts$, and add darts 0 through 4 from $\cmap_0$ to $\NonLeafDarts$ because they are ancestors of darts we added from $\cmap_1$.
Finally, with $\ell=2$, the set of leaf elements is $\operatorname{leaf}_2=\{\operatorname{face}(60),\operatorname{face}(64),\operatorname{face}(68),\operatorname{face}(84),\operatorname{face}(88),\operatorname{face}(92)\}$.
We add darts 60-71 and 84-95 from $\cmap_2$ to $\UnrefinedLeafDarts$, add darts 38, 42, 46, 57, 60-71, 81, 84-95, 108, 112, 116 from $\cmap_2$ to $\LeafDarts$, and add darts 1 and 2 from $\cmap_0$ and darts 6, 9, and 12-15 from $\cmap_1$ to $\NonLeafDarts$ because they are ancestors of darts we added from $\cmap_2$.
Finally, we remove the contents of $\NonLeafDarts$ from $\LeafDarts$, which contents are darts 0 through 4 from $\cmap_0$ and darts 6, 9, and 12-15 from $\cmap_1$.
This gives a final set of 46 darts in $\LeafDarts$, which are those that are shown in the final $\cmap_h$ with their indices mapped to the hierarchical dart space using $\mathsf{d}_h=\operatorname{hierDart}(\mathsf{d})$.

After finding all the leaf darts, we need to compute any $\phi_1$ and $\phi_{-1}$ operations that differ from the $\phi$ operations of the darts on their respective refinement levels.
For this example, there are three faces that need these explicit phi operations stored.
We begin \cref{alg:hierphi1} at line 15 with $\ell=1$.
In two dimensions, line 16 just loops through the leaf faces.
In this level, those are $\operatorname{leaf}_1=\{\operatorname{face}(0),\operatorname{face}(4),\operatorname{face}(8)\}$, as mentioned above.
On $\operatorname{face}(0)$ we see that all the darts of the face are included in $\cmap_h$, so we skip the face on line 18 of the algorithm.
On $\operatorname{face}(4)$, dart 6 is not included in $\cmap_h$, so we compute the $\textproc{phiChain}$ of the face, looping around the darts of the face in $\phi_1$ order, and adding all their leaf descendants, where a dart may be its own descendant.
For this face, the $\phi_1$ chain is darts 4 and 5 from $\cmap_1$, then darts 46, 42, and 38 from $\cmap_2$, then dart 7 from $\cmap_1$.
The process is similar for $\operatorname{face}(8)$, which has dart 9 which is not included in $\cmap_h$.
We then proceed to $\ell=0$, which has only one leaf element, $\operatorname{face}(4)$.
The darts of this face are not all in $\mathbf{B}_L$; dart 4 has leaf descendants.
The $\textproc{phiChain}$ procedure applied to this face gives a dart chain of dart 16 from $\cmap_1$, then darts 108, 112, and 116 from $\cmap_2$, then darts 5, 6, and 7 from $\cmap_0$.
Each of these chains is used to create the $\phi_1$ and $\phi_{-1}$ mapping $\mathbf{M}$ by adding each item in the chain pointing to the next item in the chain (line 12), and also the last pointing to the first (line 10).

\subsubsection{Adaptive Refinement}

Many applications for hierarchical B-splines require adaptive refinement, where the domains $\Omega_i$ are iteratively changed based on an error metric or other flagging method to converge to a solution of a desired accuracy.
\Cref{alg:hierleafdarts,alg:hierphi1} need not be run completely from scratch in this scenario.
For \cref{alg:hierleafdarts} we remove from $\UnrefinedLeafDarts$ all the darts of elements that were leaf elements but now are not, and from $\LeafDarts$ all descendants of those we removed from $\UnrefinedLeafDarts$, as well as other darts in the edges of those descendants.
After those have been removed, we can run \cref{alg:hierleafdarts} without further emptying $\LeafDarts$ and $\UnrefinedLeafDarts$, and with only the new leaf elements as input.
This update process is of linear complexity with the number of leaf elements removed plus the number of leaf elements added.
Prior to performing \cref{alg:hierphi1} we remove all darts that were removed from $\LeafDarts$ from $\mathbf{M}$ as well.
After this removal process we run \cref{alg:hierphi1} on the leaf elements that have been added and the elements adjacent to those added leaf elements, which can be found by looking for ancestors or descendants of $\phi_\Dim(\mathsf{d})$ which are in $\UnrefinedLeafDarts$, for all darts of added leaf elements.
The complexity of this update is proportional to the number of new leaf elements (or adjacent elements) which contain darts from multiple refinement levels.

\section{Evaluation}\label{sec:results}

We now discuss the scalability of \cref{alg:hierleafdarts,alg:hierphi1}, after which we show an application of the hierarchical C-map to the extraction of a watertight level set of a function in the hierarchical spline space.

\subsection{Scaling}

\Cref{alg:hierleafdarts} scales roughly as $\mathcal{O}(\numcells{\cmap_h})$.
\Cref{alg:hierphi1} scales with the number of faces in $\cmap_h$ that contain darts from multiple refinement levels (see the short circuit on line 4 of \cref{alg:hierphi1}).
Note that both the for loop on line 10 of \cref{alg:hierleafdarts} and the for loop on line 18 of \cref{alg:hierphi1} can be parallelized easily as long as appropriate care is taken while adding entries to $\NonLeafDarts$, $\UnrefinedLeafDarts$, $\LeafDarts$, and $\mathbf{M}$.

We describe here a series of example refinements and list their timing on a single thread of an Apple M3 Max MacBook Pro with 48 GB of RAM.
In an attempt to report the timing when there is the least interruption by the CPU scheduler, we run each case multiple times and take the lowest recorded timing of the tries.

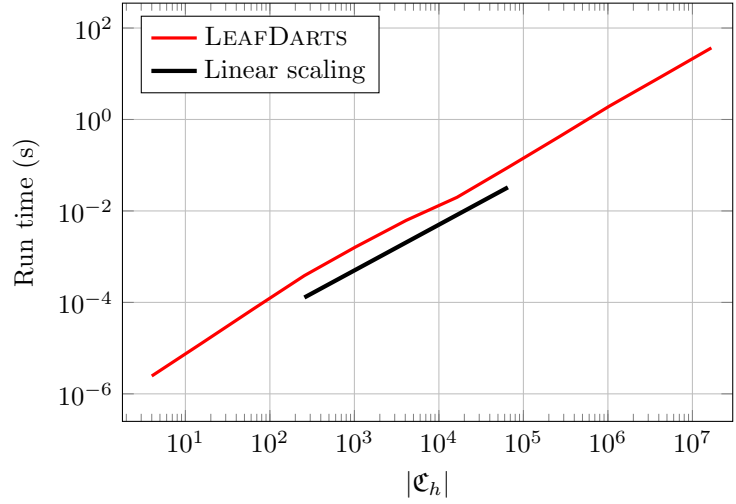
\begin{figure}
    \centering
    \begin{tikzpicture}
        \begin{loglogaxis}[
            width=3.8in,
            height=2.8in,
            xlabel={$\numcells{\cmap_h}$},
            ylabel={Run time (s)},
            grid=major,
            legend pos=north west,
            legend cell align=left,
            ytick={1e-6,1e-4,1e-2,1,1e2},
            ymin=2.5e-7,
            ymax=350,
            xmin=1.8,
            xmax=3e7
        ]

        \addplot[red, solid, very thick] coordinates {
            (4, 2.458e-06)
            (16, 1.3209e-05)
            (64, 7.2209e-05)
            (256, 0.000384917)
            (1024, 0.00161629)
            (4096, 0.00620988)
            (16384, 0.0198515)
            (65536, 0.0891848)
            (262144, 0.413752)
            (1048576, 1.97378)
            (4194304, 8.45889)
            (16777216, 36.7706)
        };
        \addlegendentry{\textproc{LeafDarts}}


        \addplot[black, solid, ultra thick] coordinates {
            (256, 0.000128)
            (1024, 0.000512)
            (4096, 0.002048)
            (16384, 0.008192)
            (65536, 0.032768)
        };
        \addlegendentry{Linear scaling}

        \end{loglogaxis}
    \end{tikzpicture}
    \caption{Scaling of \cref{alg:hierleafdarts} with respect to number of leaf elements for global refinement, i.e., $\operatorname{leaf}_{\NumRefinementLevels-1}$ is all the elements of $\cmap_{\NumRefinementLevels-1}$, and all other $\operatorname{leaf}_i$ are empty sets.}
    \label{fig:global_timing}
\end{figure}

First we simply do global refinement using the hierarchical algorithms.
We begin with a two-dimensional tensor product combinatorial map with one element, and dyadically refine it repeatedly to $\NumRefinementLevels$ levels.
All sets of leaf elements are left empty except for $\operatorname{leaf}_{\NumRefinementLevels-1}$, which contains all the elements in $\cmap_{\NumRefinementLevels-1}$.
Timing of \cref{alg:hierleafdarts} as $\NumRefinementLevels$ is varied from two to 14 levels---and $\numcells{\cmap_h}$ scales from $2^2$ to $2^{26}$---are shown in \cref{fig:global_timing}, with $\numcells{\cmap_h}$ as the x-axis.
We see that \cref{alg:hierleafdarts} scales roughly linearly with $\numcells{\cmap_h}$, as expected.
For this setup \cref{alg:hierphi1} is not shown, but is doing almost no work, simply checking that all darts of cells in $\operatorname{leaf}_{\NumRefinementLevels-1}$ are contained in $\LeafDarts$.
This action also scales linearly with $\numcells{\cmap_h}$, but with a much smaller constant, and in each case tested for \cref{fig:global_timing} the timing is less than 10~ns.

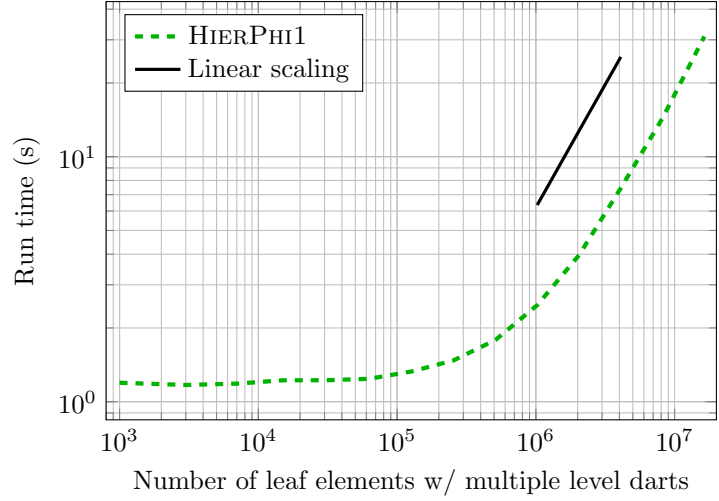
\begin{figure}
    \centering
    \begin{tikzpicture}
        \begin{loglogaxis}[
            width=3.8in,
            height=2.8in,
            xlabel={Number of leaf elements w/ multiple level darts},
            ylabel={Run time (s)},
            grid=both,
            legend pos=north west,
            legend cell align=left,
            xmin=800,
            xmax=2e7
        ]

        \addplot[green!70!black, dashed, ultra thick] coordinates {
            (1000, 1.19488)
            (3000, 1.17036)
            (7000, 1.18518)
            (15000, 1.22288)
            (31000, 1.2235)
            (63000, 1.2405)
            (127000, 1.32854)
            (255000, 1.47429)
            (511000, 1.78595)
            (1023000, 2.47806)
            (2047000, 3.96645)
            (4095000, 7.44809)
            (8191000, 14.3904)
            (16383000, 30.9645)
        };
        \addlegendentry{\textproc{HierPhi1}}

        \addplot[black, solid, very thick] coordinates {
            (1023000, 6.35)
            (2047000, 12.75)
            (4095000, 25.55)
        };
        \addlegendentry{Linear scaling}

        \end{loglogaxis}
    \end{tikzpicture}
    \caption{Scaling of \cref{alg:hierphi1} with respect to number of elements in $\cmap_h$ which contain darts from multiple refinement levels.}
    \label{fig:phi1_timing}
\end{figure}

Our next benchmark focuses on the scaling of \cref{alg:hierphi1}.
The main work in this algorithm is done the number of times that there are elements in $\cmap_h$ that are not skipped by the statement on line 18 of \cref{alg:hierphi1}.
To see this scaling, we devise a series of $\cmap_h$ that have the same number of elements, but differing numbers of elements which contain darts from multiple refinement levels.
We do this by creating $\cmap_0$ as a two-dimensional tensor product combinatorial map that is 1000 elements wide and $2^{15}$ elements tall, and $\cmap_1$ as a dyadic refinement of $\cmap_0$ that is 2000 elements wide and $2^{16}$ elements tall.
We run a series of tests, varying a parameter $n_\mathrm{groups}$ between values 2 and $2^{14}$, which is used in the following way:
We form the $\operatorname{leaf}_i$ by dividing the $2^{15}$ columns of $\cmap_0$ into $n_\mathrm{groups}$ equal contiguous groups, and the elements from the odd-numbered groups are added to $\operatorname{leaf}_0$, while the elements from the even-numbered groups have their children added to $\operatorname{leaf}_1$.
This results in the number of elements with multiple dart levels ranging from 1000 to 16,383,000.
We can see in \cref{fig:phi1_timing} that as the number of cells with darts from multiple levels increases, the timing of \cref{alg:hierphi1} asymptotically approaches linear scaling with that number.
The timing of \cref{alg:hierleafdarts} is not shown in the figure, but is constant across all the runs shown here, as the number of elements in $\cmap_h$ does not change.

\subsection{Application to Level Set Extraction}

\begin{figure}
    \centering
    \includegraphics[width=0.6\textwidth]{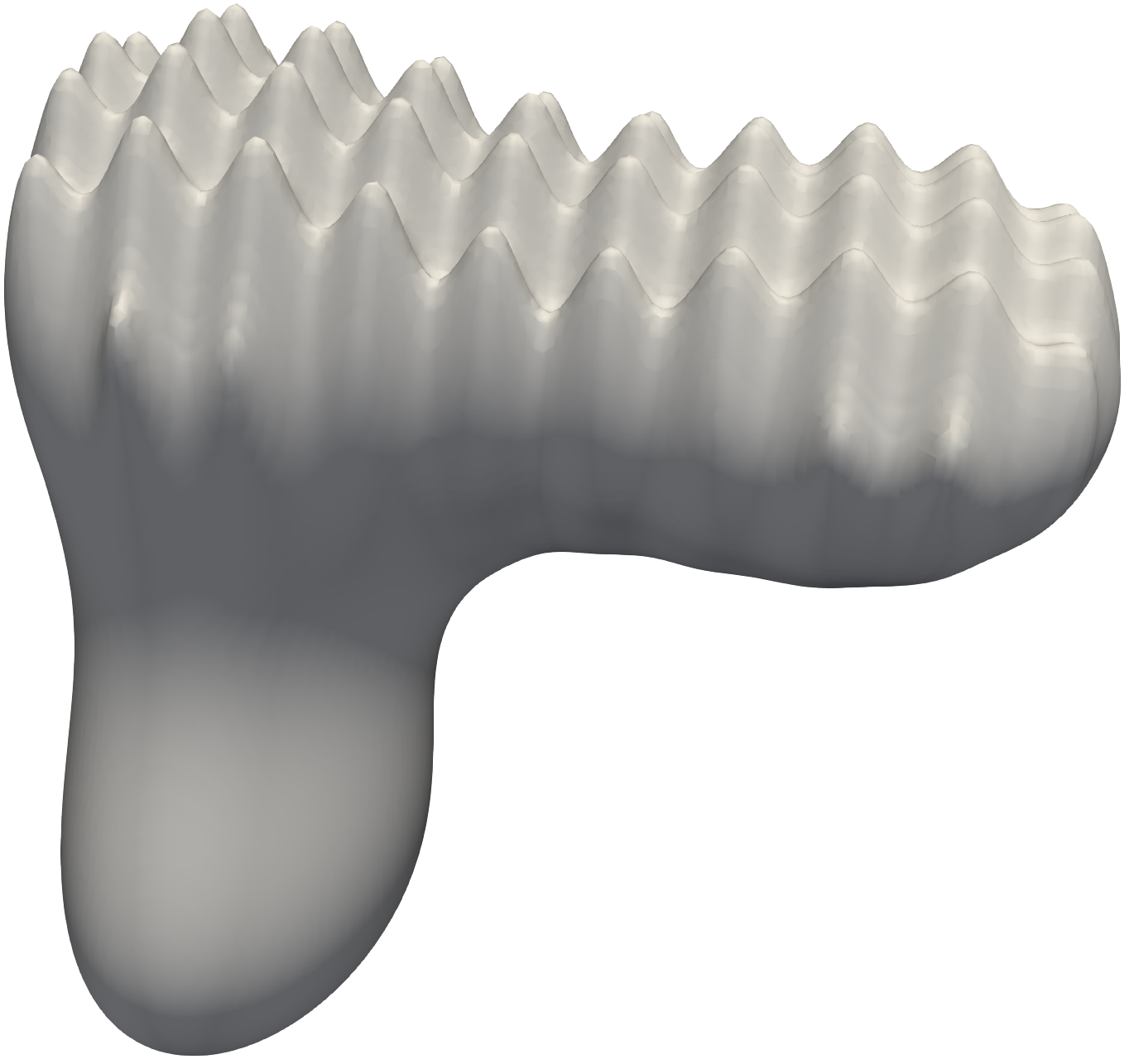}
    \caption{Level set $f^{-1}(0.15)$, which we attempt to recover.}
    \label{fig:levelset}
\end{figure}

\begin{figure}
    \centering
    \includegraphics[width=0.6\textwidth]{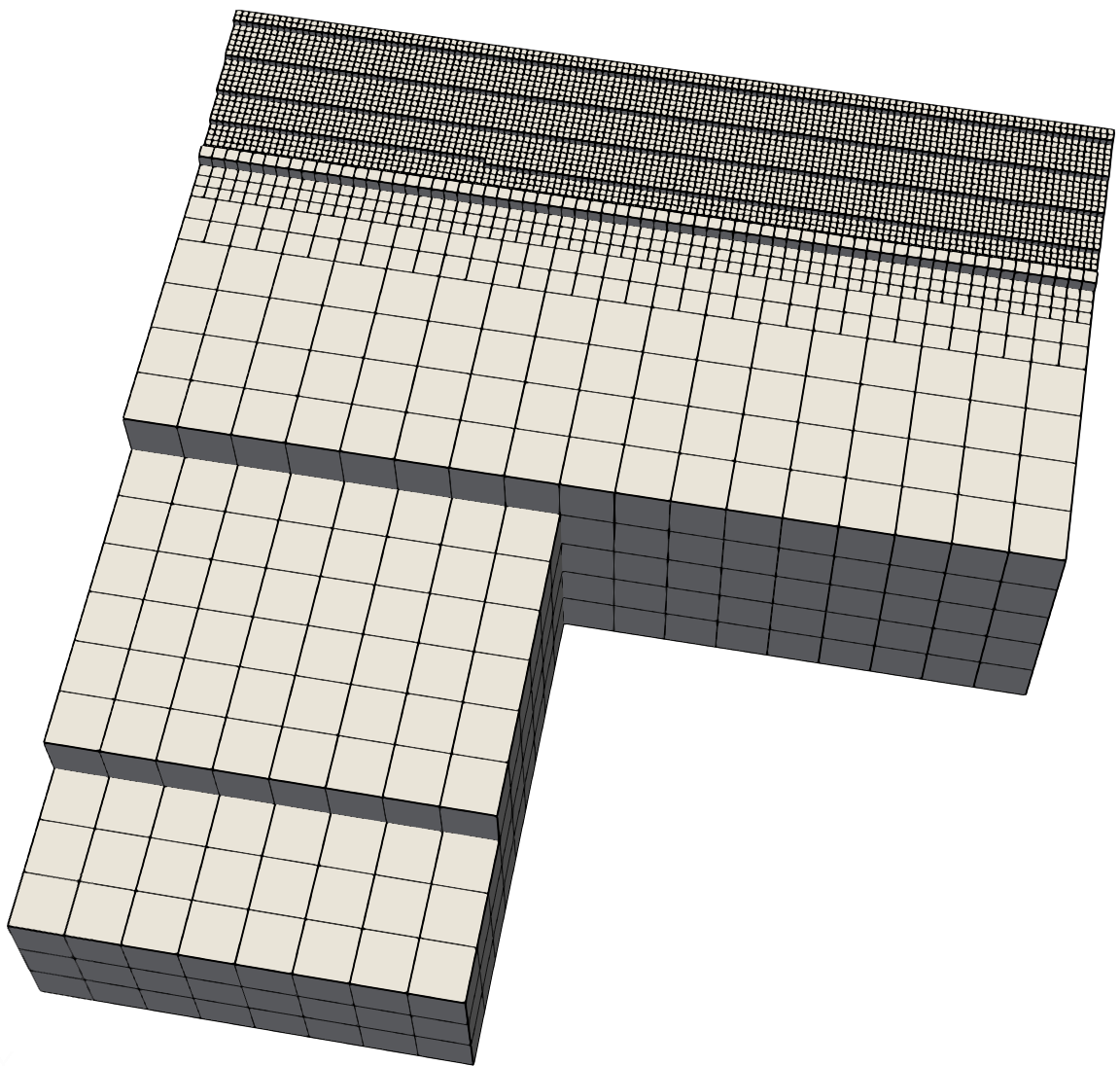}
    \caption{A cutaway of the hierarchical B\'ezier mesh used to compute the level set from \cref{fig:levelset}.}
    \label{fig:levelset_mesh}
\end{figure}

As an application, we compute the level set of a function defined on a three-dimensional truncated hierarchical B-spline basis.
This is a typical operation in applications such as topology optimization \cite{Noel:2020,Schmidt:2025}, multiphase flow \cite{Yan:2019}, among others.
We simulate the process of retrieving the level set of a computed function, and instead simply project the following function into the B-spline space:
\begin{align*}
    f(x,y,z)=&\\
    &x^2(x-1)^2+y^2(y-1)^2+x^2y^2+z^2\\
    &+\frac{(x-1)^6}{100}(\sin{40z}+\sin{40y}).
\end{align*}
The 0.15 level set of this function, which is the level set we will extract, is shown in \cref{fig:levelset}.
The unrefined B-spline space is a 3-patch B-spline with the patch regions defined as
\begin{align*}
    \Omega^0&=[-0.45,0.45]\times[-0.35,0.45]\times[-0.40,0.46]\\
    \Omega^1&=[0.45,1.35]\times[-0.35,0.45]\times[-0.40,0.46]\\
    \Omega^2&=[-0.45,0.45]\times[0.45,1.35]\times[-0.40,0.46],
\end{align*}
with those regions chosen to contain the 0.15 level set of $f$.
We use cubic splines with uniform knot vectors using element sizes of 0.1 in the $x$ and $y$ directions, and 0.1075 in the $z$ direction.
Refinement is chosen so that $\Omega^1$ is unrefined, i.e. $\Omega^1\cap\Omega_1=\emptyset$, and as $x$ decreases in $\Omega^0$ and $\Omega^2$ the refinement level around the 0.15 level set increases.
\Cref{fig:levelset_mesh} shows a cross section of the hierarchical mesh, where the refinement pattern is evident.

For the purposes of this comparison, we will extract an approximate level set by making the simplifying assumption that the function is trilinear on each element.
We assume that each face of the mesh is intersected at most once by the level set.
As a further approximation to the true level set, we find the intersections of each edge on the level set under the trilinear assumption, and linearly triangulate between the intersections.

The typical approach to defining hierarchical spline bases, with the set of leaf elements being the only topological information stored, allows for a marching cubes approach to extracting this approximate level set.
This means that for each element, the function and the mesh positions are evaluated at each of the eight corners of the element, and a template is added to the surface triangulation based on which of the corners are greater or less than the desired level \cite{Lewiner:2003}.
This is doable, but results in many duplicate evaluations of the basis and also means that the resultant level set is not watertight, it is simply a triangle soup.

In contrast, with the topological information afforded by the hierarchical combinatorial map, the function value at each vertex can be calculated only once and used again every time it is needed.
This is nearly an eight times reduction in evaluations for a fully structured mesh.
Because we have the element to element adjacency information available in the combinatorial map, we can also compute a watertight representation of the level set.
The presence of T-junctions in the hierarchical mesh prevents the use of standard marching cubes templates; instead, we find all the edges that intersect the level set and make those the vertices of the level set mesh, find faces adjacent to intersected edges and make those edges of the level set mesh, and define the faces of the level set by their boundary edges.
This is enough to define a polygonal mesh representing the level set.

Calculation of the positions of the vertices under the isoparametric approach to defining the geometry also involves evaluating the spline basis, but in both approaches this is only done for the vertices adjacent to the subset of cells that intersect the level set---for marching cubes, this is every volumetric cell that intersects the level set, and for the topological approach this is every edge that intersects the level set.

An additional optimization is available because of the topological information: if it is known that the level set is one connected component, or if only a single connected component of the level set is desired, we can flood only to neighboring elements of those that intersect rather than checking every edge of the whole mesh.
We include this method as well in the following comparison.

\begin{figure}
    \centering
    \includegraphics[width=0.6\textwidth]{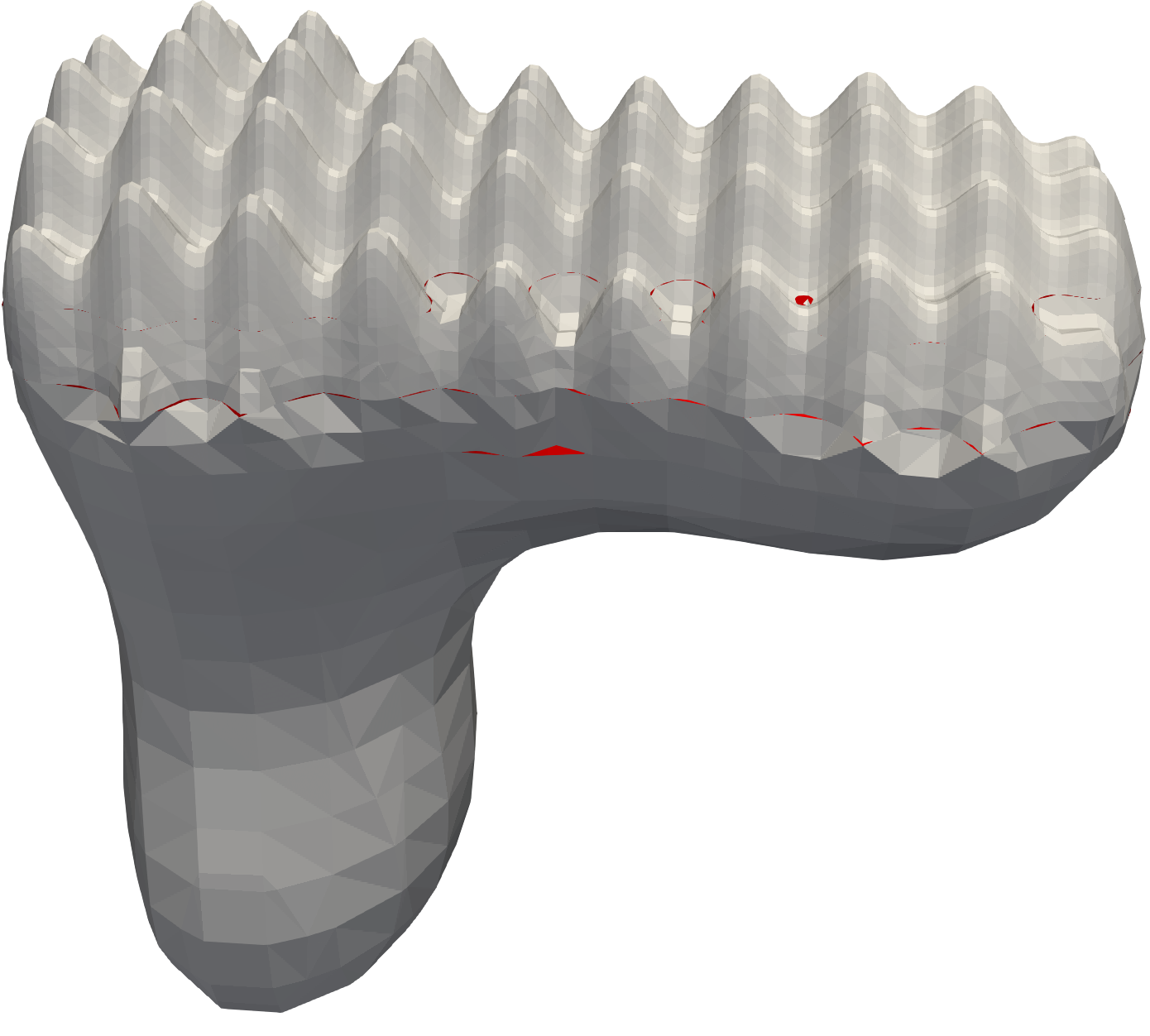}
    \caption{The level set computed using marching cubes.  The inside of the surface is colored red to show holes in the surface, which are due to the inability of marching cubes to deal with hierarchical meshes.}
    \label{fig:marching_cubes_levelset}
\end{figure}

\begin{figure}
    \centering
    \includegraphics[width=0.6\textwidth]{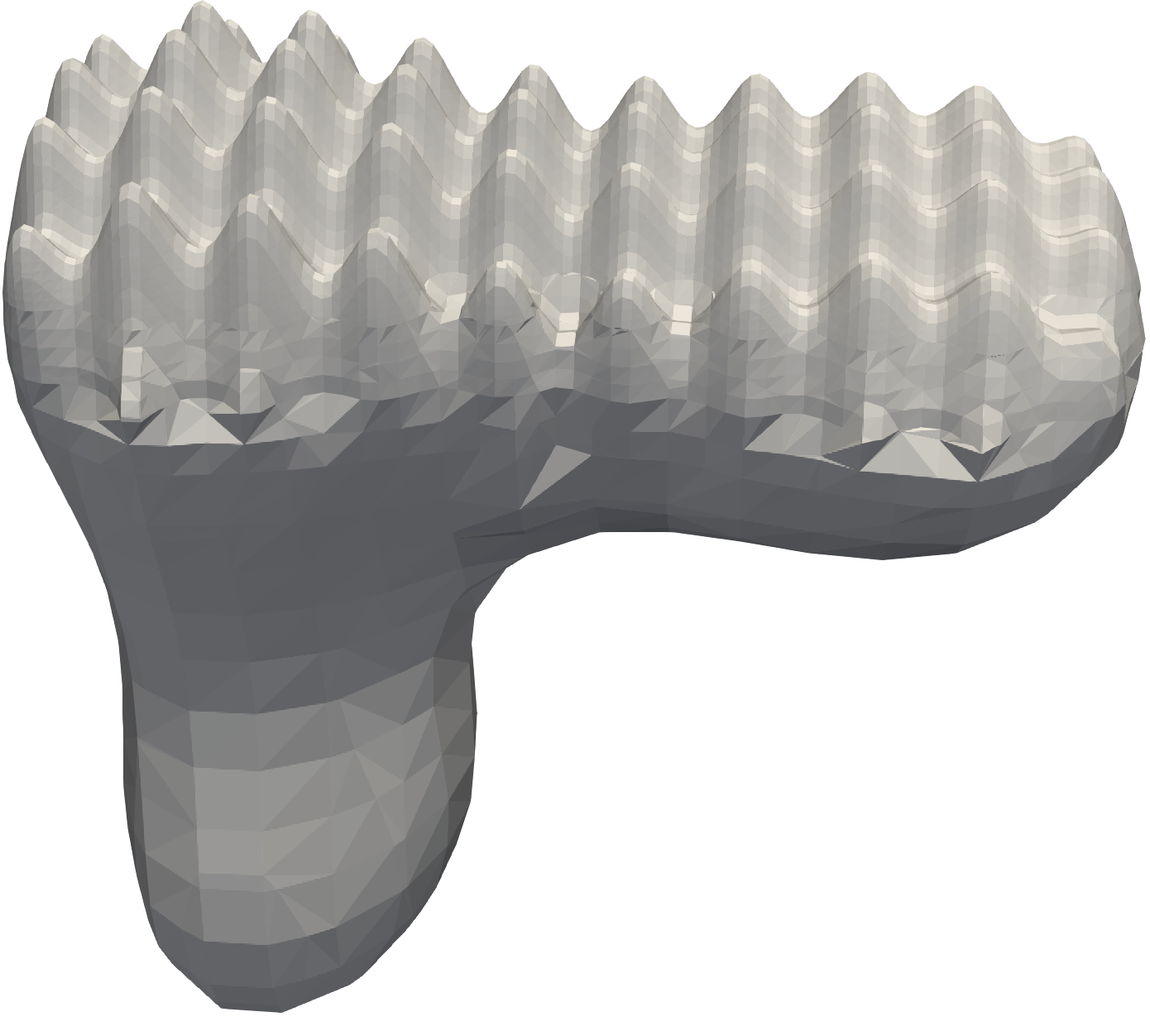}
    \caption{The level set computed using topological information. The resulting surface mesh is watertight.}
    \label{fig:cmap_method_levelset}
\end{figure}

The surface reconstructed by the marching cubes algorithm is shown in \cref{fig:marching_cubes_levelset}.
It has gaps where there are transitions between coarser and finer elements, shown by the colored interior of the surface that can be seen through the gaps.
In contrast, \cref{fig:cmap_method_levelset} shows the level set computed with the topological information from the combinatorial map.
There are no gaps in this level set; it is a watertight polygonal surface mesh.
The surface here is the same whether we use the local flood optimization or not.
We summarize the cost and benefits of the three methods in \cref{tab:levelset_compare}.
Note that the presence of the topological information causes a 7.5 times reduction in the number of function evaluations required to construct the level set.
It also reduces the number of vertex position evaluations needed by a factor of 3.95.
If we use the local flood optimization, the number of function evaluations goes down by another factor of 6.7, for a total reduction factor compared to marching cubes of more than 50 times.

\begin{table*}
    \centering
    \begin{tabular}{ccccc}
        \toprule
        Approach & Watertight & \makecell{\# of $f$\\evaluations} & \makecell{\# of position\\evaluations} & Time (s) \\
        \midrule
        Marching Cubes & \XSolidBrush & 1,758,464 & 134,808 & 74.4\\
        $\cmap_h$ full flood & \Checkmark & 234,300 & 34,154 & 20.3 \\
        $\cmap_h$ local flood & \Checkmark & 34,890 & 34,154 & 5.4 \\
        \bottomrule
    \end{tabular}
    \caption{Comparison of marching cubes approach, an approach with $\cmap_h$ that checks the entire mesh, and an approach with $\cmap_h$ that uses adjacency information to only check part of the mesh.}
    \label{tab:levelset_compare}
\end{table*}

For the example shown here, run on a single thread on the same machine from the previous section, this leads to the local flood approach taking less than one tenth the time of the marching cubes approach, and the global topologically informed approach taking less than a third of the time of marching cubes, where we are using a known performant implementation of marching cubes (namely the implementation of traditional marching cubes at \cite{lewinermarchingcubes}), and the topological algorithm has not had extensive optimization for speed.

\section{Conclusions}\label{sec:conclusions}

We have shown here construction algorithms for a combinatorial map representing the B\'ezier mesh of hierarchical B-splines and other hierarchical splines.
The benefits of combinatorial maps have been shown in the literature, but a combinatorial map construction routine for hierarchical splines has not previously been published, to the knowledge of the authors.
Our algorithm works directly from the input typically used to define the refinement domains in hierarchical splines, namely the leaf elements from each refinement level.
Because we use implicitly defined refinement levels, the total memory usage is reasonable compared to state of the art combinatorial maps, and the structure of the map we propose can be transferred to other representations if needed after construction.
There are also ways to update the combinatorial map in an adaptive refinement setting without starting from scratch.
We've shown the scaling of the construction algorithms, and also the use of the resulting combinatorial map for the extraction of a watertight level set of a function in the spline space.

This algorithm is particularly appropriate for the case where you can represent the global refinement levels implicitly.
Indeed, if you must represent the refinement levels explicitly, the memory use of the data structure would certainly be worse than that of a globally refined mesh.
Despite this limitation, there is potential for these algorithms to be employed in other hierarchical splines that do not use only cubical cells if such an implicit representation of the refinement level combinatorial maps is available.

\section*{Acknowledgements}
	C. Goates and K. Shepherd were partially supported by National Science Foundation, United States under Grant No. 2245491.  
	Any opinion, findings, and conclusions or recommendations expressed in this material are those of the authors and do not necessarily reflect the views of the National Science Foundation.

\bibliography{references}

\begin{appendices}
\section{Failure of the algorithms on a non-conforming B\'ezier mesh}\label{sec:tjunctionfail}

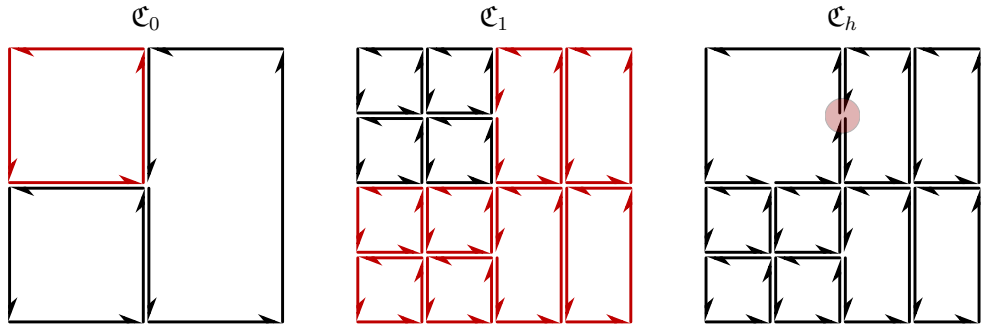
\begin{figure}
\centering
\def\EPS{0.04}

\tikzset{dart/.style={-{Stealth[harpoon]}, very thick, line cap=round, line join=round}}

\newcommand{\dartcell}[5][]{%
  \coordinate (dA) at ({#2+\EPS}, {#3+\EPS});
  \coordinate (dB) at ({#4-\EPS}, {#3+\EPS});
  \coordinate (dC) at ({#4-\EPS}, {#5-\EPS});
  \coordinate (dD) at ({#2+\EPS}, {#5-\EPS});
  \draw[dart,#1] (dA) -- (dB);
  \draw[dart,#1] (dB) -- (dC);
  \draw[dart,#1] (dC) -- (dD);
  \draw[dart,#1] (dD) -- (dA);
}

\newcommand{\dartcellSplitLeft}[6][]{%
  \coordinate (dA)  at ({#2+\EPS}, {#3+\EPS});
  \coordinate (dB)  at ({#4-\EPS}, {#3+\EPS});
  \coordinate (dC)  at ({#4-\EPS}, {#5-\EPS});
  \coordinate (dD)  at ({#2+\EPS}, {#5-\EPS});
  \coordinate (dDm) at ({#2+\EPS}, {#6+\EPS});
  \coordinate (dDm2) at ({#2+\EPS}, {#6-\EPS});
  \draw[dart,#1] (dA)  -- (dB);
  \draw[dart,#1] (dB)  -- (dC);
  \draw[dart,#1] (dC)  -- (dD);
  \draw[dart,#1] (dD)  -- (dDm);
  \draw[dart,#1] (dDm2) -- (dA);
}

\newcommand{\dartcellSplitRight}[6][]{%
  \coordinate (dA)  at ({#2+\EPS}, {#3+\EPS});
  \coordinate (dB)  at ({#4-\EPS}, {#3+\EPS});
  \coordinate (dC)  at ({#4-\EPS}, {#5-\EPS});
  \coordinate (dD)  at ({#2+\EPS}, {#5-\EPS});
  \coordinate (dBm) at ({#4-\EPS}, {#6-\EPS});
  \coordinate (dBm2) at ({#4-\EPS}, {#6+\EPS});
  \draw[dart,#1] (dA)  -- (dB);
  \draw[dart,#1] (dB)  -- (dBm);
  \draw[dart,#1] (dBm2) -- (dC);
  \draw[dart,#1] (dC)  -- (dD);
  \draw[dart,#1] (dD)  -- (dA);
}

\newcommand{\dartcellSplitBottom}[6][]{%
  \coordinate (dA)  at ({#2+\EPS}, {#3+\EPS});
  \coordinate (dB)  at ({#4-\EPS}, {#3+\EPS});
  \coordinate (dC)  at ({#4-\EPS}, {#5-\EPS});
  \coordinate (dD)  at ({#2+\EPS}, {#5-\EPS});
  \coordinate (dAm) at ({#6-\EPS}, {#3+\EPS});
  \coordinate (dAm2) at ({#6+\EPS}, {#3+\EPS});
  \draw[dart,#1] (dA)  -- (dAm);
  \draw[dart,#1] (dAm2) -- (dB);
  \draw[dart,#1] (dB)  -- (dC);
  \draw[dart,#1] (dC)  -- (dD);
  \draw[dart,#1] (dD)  -- (dA);
}

\newcommand{\dartcellSplitRightBottom}[7][]{%
  \coordinate (dA)  at ({#2+\EPS}, {#3+\EPS});
  \coordinate (dB)  at ({#4-\EPS}, {#3+\EPS});
  \coordinate (dC)  at ({#4-\EPS}, {#5-\EPS});
  \coordinate (dD)  at ({#2+\EPS}, {#5-\EPS});
  \coordinate (dBm) at ({#4-\EPS}, {#6-\EPS});
  \coordinate (dBm2) at ({#4-\EPS}, {#6+\EPS});
  \coordinate (dAm) at ({#7-\EPS}, {#3+\EPS});
  \coordinate (dAm2) at ({#7+\EPS}, {#3+\EPS});
  \draw[dart,#1] (dA)  -- (dAm);
  \draw[dart,#1] (dAm2) -- (dB);
  \draw[dart,#1] (dB)  -- (dBm);
  \draw[dart,#1] (dBm2) -- (dC);
  \draw[dart,#1] (dC)  -- (dD);
  \draw[dart,#1] (dD)  -- (dA);
}
\resizebox{\textwidth}{!}{
\begin{tikzpicture}[x=1cm, y=1cm]

\begin{scope}[shift={(0,0)}]
  \node[above, font=\large] at (2, 4.1) {$\cmap_0$};

  \dartcell{0}{0}{2}{2}
  \dartcell[red!75!black]{0}{2}{2}{4}

  \dartcellSplitLeft{2}{0}{4}{4}{2.0}
\end{scope}

\begin{scope}[shift={(5,0)}]
  \node[above, font=\large] at (2.0, 4.1) {$\cmap_1$};

  \foreach \j in {0,1}{
    \foreach \i in {0,1}{
      \pgfmathsetmacro{\llx}{\i*1.0}
      \pgfmathsetmacro{\lly}{\j*1.0}
      \dartcell[red!75!black]{\llx}{\lly}{\llx+1.0}{\lly+1.0}
    }
  }
  \foreach \j in {2,3}{
    \foreach \i in {0,1}{
      \pgfmathsetmacro{\llx}{\i*1.0}
      \pgfmathsetmacro{\lly}{\j*1.0}
      \dartcell{\llx}{\lly}{\llx+1.0}{\lly+1.0}
    }
  }

  \dartcellSplitLeft[red!75!black]{2.0}{0.0}{3.0}{2.0}{1.0}
  \dartcell[red!75!black]{3.0}{0.0}{4.0}{2.0}
  \dartcellSplitLeft[red!75!black]{2.0}{2.0}{3.0}{4.0}{3.0}
  \dartcell[red!75!black]{3.0}{2.0}{4.0}{4.0}
\end{scope}

\begin{scope}[shift={(10,0)}]
  \node[above, font=\large] at (2.0, 4.1) {$\cmap_h$};

  \dartcellSplitRightBottom{0.0}{2.0}{2.0}{4.0}{3.0}{1.0}

  \foreach \j in {0,1}{
    \foreach \i in {0,1}{
      \pgfmathsetmacro{\llx}{\i*1.0}
      \pgfmathsetmacro{\lly}{\j*1.0}
      \dartcell{\llx}{\lly}{\llx+1.0}{\lly+1.0}
    }
  }

  \dartcellSplitLeft{2.0}{0.0}{3.0}{2.0}{1.0}
  \dartcell{3.0}{0.0}{4.0}{2.0}

  \dartcellSplitLeft{2.0}{2.0}{3.0}{4.0}{3.0}
  \dartcell{3.0}{2.0}{4.0}{4.0}

\draw[fill=red!60!black, line width=0pt, opacity=0.3]
  (2,3) circle[radius=0.25];
\end{scope}

\end{tikzpicture}
}
\caption{Failure of \cref{alg:hierleafdarts} to create the correct topology when a T-junction is present in the B\'ezier mesh of a refinement level.}\label{fig:t_junction_failure}
\end{figure}

\Cref{alg:hierleafdarts,alg:hierphi1} may not create the correct topology when the $\cmap_i$ do not have conforming B\'ezier meshes.
We show in \cref{fig:t_junction_failure} an example of a case where an incorrect topology is created.
This example has two refinement levels; the first is a simple T-junction with three faces in the mesh, and the second is a dyadic refinement of the first.
This second refinement level contains two T-junctions just as a consequence of the dyadic refinement.
The leaf elements from $\cmap_0$ and $\cmap_1$ have their darts colored red.

\Cref{alg:hierleafdarts} marks six darts from $\cmap_0$ as leaf: the four in red, and two that are the $\phi_2$ of red darts.
Then it marks all the red darts from $\cmap_1$ as leaf, and four additional darts that are the $\phi_2$ of red darts from that level.
Finally, it unmarks four darts from $\cmap_0$ as leaf, because their descendants were made leaf---namely, the two black darts from level 0 that were marked as leaf and their $\phi_2$ darts.

The final combinatorial map is shown on the right.
As highlighted in red, it contains a hanging vertex that does not represent any vertex in the hierarchical B\'ezier mesh.

\end{appendices}
\end{document}